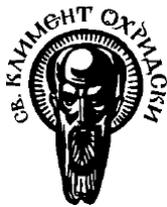

Софийски Университет "Св. Климент Охридски"
Факултет по Математика и Информатика
Катедра "Информационни технологии"
Специализация „Био- и медицинска информатика"

# ДИПЛОМНА РАБОТА

Тема:

Визуален подход за извличане на закономерности от бази от данни с медицинска информация чрез използване на FastMap алгоритъм

Дипломант: Петър Стефанов Кормушев, Ф№ М-21546

Научен ръководител: доц. д-р Антоний Попов
ФМИ, СУ "Св. Климент Охридски"

"Data is not information, Information is not knowledge, Knowledge is not understanding, Understanding is not wisdom."

Cliff Stoll & Gary Schubert

———————————————

"Данните не са информация, Информацията не е знание, Знанието не е разбиране, Разбирането не е Мъдрост."

Клиф Стол & Гари Шуберт

# Резюме

Бързият прогрес в развитието на средства за придобиване и съхраняване на информация доведе до натрупването на големи медицински бази от данни. Това огромно количество събрани данни значително превишава възможностите на човек те да бъдат ефективно използвани без помощта на специализирани мощни средства за анализ. Като резултат, данните, събрани в огромните бази, са се превърнали в "гробници от архиви", които почти не са посещавани от никого. Така се стига до ситуация, описвана като "богата на данни, но бедна на информация" [Агре, 2003].

За запълване на тази все по-разширяваща се пропаст между данните и информацията се прилагат различни подходи от научната дисциплина "Извличане на закономерности от данни" (Data Mining), която цели да превърне тези гробниците в "златни мини", от които се добиват знания. Извличането на закономерности от данни (ИЗД) се състои в анализ на (често много големи) множества от наблюдавани данни с цел да бъдат открити в тях неочаквани зависимости или те да бъдат обобщени и представени по нови начини, които са разбираеми и полезни за притежателите им. Един от възможните подходи за откриване на зависимости е визуалният, при който данните се подлагат на обработка и се визуализират по някакъв обобщен начин, позволяващ анализирането им от експерти в областта.

Целта на настоящата дипломна работа е проектиране и създаване на софтуерна система за визуализация на многомерни, класифицирани медицински данни. Това се постига чрез прилагане на FastMap алгоритъм за намаляване размерността на данните, така че да стане възможно тяхното двумерно визуализиране. Софтуерната система е разделена на два отделни инструмента, които ще наричаме „Конструктор" и „Визуализатор", като изборът на тези имена ще стане ясен в хода на изложението.

Изложението е разделено на четири големи глави. В първа глава се прави въведение в предметната област и се дефинира по-точно целта на дипломната работа. Във втора глава се прави анализ на изискванията за създаване на извадки от големи бази от данни и се проектира инструментът „Конструктор". В трета глава се прави описание на алгоритъма FastMap за намаляване размерността на данни и се проектира инструментът „Визуализатор". В четвърта глава се описва конкретната реализация на двата инструмента както на ниво потребителски интерфейс, така и на ниво вътрешна реализация. Изложението завършва с обобщение и насоки за бъдещо развитие.

# Благодарности

# Съдържание

# 1. Въведение

Бързият прогрес в развитието на средства за придобиване и съхраняване на информация, наблюдаван в последните години, доведе до натрупване на големи бази от данни в практически всички области на човешката дейност. В медицината това е особено осезаемо, тъй като медицинските бази включват изключително разнообразни данни за пациентите – информация за извършените прегледи, издадени направления, антропометрични данни, активни диспансеризации, двумерни снимки от рентгенография, тримерни снимки от компютърна томография, резултати от назначени медицински изследвания, ДНК проби и т.н.

Това огромно количество събирани данни значително превишава възможностите на човек те да бъдат ефективно използвани без помощта на специализирани мощни средства за анализ. Като резултат, данните, събрани в огромните бази от данни, са се превърнали в "гробници от архиви", които почти не са посещавани от никого. Така се стига до ситуация, описвана като "богата на данни, но бедна на информация" [Agre, 2003].

За запълване на тази все по-разширяваща се пропаст между данните и информацията се прилагат различни подходи от научната дисциплина "Извличане на закономерности от данни" (Data Mining), която цели да превърне тези гробниците в "златни мини", от които се добиват знания.

## *1.1. Извличане на закономерности от данни*

Извличането на закономерности от данни (ИЗД) се състои в анализ на (често много големи) множества от *наблюдавани данни* с цел да бъдат открити в тях неочаквани зависимости, или те да бъдат обобщени и представени по нови начини, които са разбираеми и полезни за притежателите на данните.

Изведените от данните зависимости и обобщения често се наричат модели или шаблони, които се представят чрез линейни уравнения, правила, клъстери, графи, дървета и т.н. Изразът "*наблюдавани данни*" е използван в по-горната дефиниция като противоположност на "*експериментални данни*". Обикновено ИЗД работи с данни, които са били събрани за цели, различни от ИЗД-анализ, което означава, че поставените от ИЗД цели не оказват никакво влияние на стратегията за събиране на данни. Това е една от характеристиките, по които ИЗД значително се отличава от статистиката, където често данните се събират чрез използване на ефективни стратегии за получаване на отговори на специфични въпроси. По тази причина ИЗД често се нарича "вторичен" анализ на данни.

В дефиницията на ИЗД се набляга на големия размер на базите от данни. При работа с малки множества от данни могат да бъдат използвани методи на класическия изследователски анализ, прилаган от статистиците. В случая на големи бази от данни възникват нови проблеми. Някои от тях са свързани с начина за ефективно съхраняване и намиране на данни, а други се отнасят за такива фундаментални въпроси, като как да бъдат избрани характерни представители на данните, как тези данни могат да бъдат проанализирани за приемливо време, как да се разбере, дали някоя намерена зависимост отразява действителната реалност, а не е само резултат от някакво случайно съвпадение само в определена част от данните и т.н.

Важната характеристика на извлечените зависимости и структури е тяхната *новост*. Ясно е, че степента на „новото" трябва да се измерва относно априорните, базови знания на потребителя. За съжаление, много малко ИЗД алгоритми взимат под внимание подобни знания. Макар че степента на новост е едно важно свойство на зависимостите, които търсим в данните, само то не е достатъчно да квалифицира една зависимост като *знание*. За целта тя трябва да бъде и разбираема.

За работещите в медицинската сфера – общопрактикуващи лекари и специалисти – един такъв по-разбираем и полезен начин за представяне на данни е графичният. За съжаление, огромните медицински бази от данни са пълни с невизуална (числова и текстова) информация, която е трудно да бъде показана в съкратен вид достатъчно разбираемо. За да се създаде подходящо графично представяне на данните, важно е да се разгледа по-внимателно естеството на данните.

## 1.2. Естество на данните

Данните представляват едно множество от измервания, направени в определена среда или върху определен процес. В най-простия случай имаме колекция от обекти, като всеки от тях се характеризира с множество от $p$ измервания, еднакви за всички обекти. В този случай такава колекция може да се разглежда като $n \times p$ матрица от данни, където $n$ редове представят $n$ обекта, върху които са били направени измервания. В различни научни дисциплини (контексти) тези обекти се наричат индивиди, случаи, единици, обекти, примери или записи.

Втората размерност на матрицата от данни съдържа резултати от $p$ измервания, направени върху всеки от обектите. Например, измерванията за медицински пациенти включват техни антропометрични данни като ръст, тегло, възраст, пол, в комбинация с други показатели като кръвно налягане, локация на болката, диагноза на заболяването и т.н. Най-често се предполага, че върху всеки обект е направено едно и също множество от измервания, макар че това не е задължително. Например, върху различни пациенти могат да бъдат извършени различни медицински тестове. Тези $p$ колонки на матрицата с данни се наричат (в различни контексти) променливи, признаци, атрибути или полета.

За да илюстрираме по-добре естеството на данните, ще разгледаме един пример. Бюрото по преброяване на населението на САЩ събира данни за населението на страната на всеки 10 години. Част от тази информация става публично достъпна, като всички данни, позволяващи конкретен индивид да бъде идентифициран, се изтриват. Данните са достъпни във вид на 5% и 1% извадки (трябва да се отбележи, че дори 1% извадка съдържа около 2.7 милиона записи). Данните съдържат десетки променливи, от типа на възраст, доход, ниво на образование и т.н. В таблица 1.1 е даден пример за подобни данни. Вижда се, че таблицата съдържа различни типове променливи – непрекъснати (например Age - възраст) и символни (например Marital Status - семейно положение). Някои стойности липсват – това е често явление в реалните бази от данни. По-сложен въпрос е наличието на шум – например, дали доходът (income) на човека с ID 248 е действително $100000 или това е грешка при въвеждането?

| ID | Age | Sex | Marital Status | Education | Income |
|---|---|---|---|---|---|
| 248 | 54 | Male | Married | High school graduate | 100000 |
| 249 | ?? | Female | Married | High school graduate | 12000 |
| 250 | 29 | Male | Married | Some college | 23000 |
| 251 | 9 | Male | Not married | Child | 0 |
| 252 | 85 | Female | Not married | High school graduate | 19798 |
| 253 | 40 | Male | Married | High school graduate | 40100 |
| 254 | 38 | Female | Not married | Less than 1st grade | 2691 |
| 255 | 7 | Male | ?? | Child | 0 |
| 256 | 49 | Male | Married | 11th grade | 30000 |
| 257 | 76 | Male | Married | Doctorate degree | 30686 |

*Таблица 1.1. Пример за данни, събирани от Бюрото по преброяване на населението на САЩ*

Различието между непрекъснати и символни променливи е важно, тъй като някои от техниките за анализ на данни, подходящи за единия тип променливи, не са подходящи за другия. Непрекъснатите променливи се измерват по числова скала и по принцип могат да приемат произволни числови стойности. Символните променливи могат да приемат само определени дискретни стойности. Символните стойности могат да бъдат подредени (т.е. да имат естествен начин на подреждане, например Education – "Степен на образование"), или номинални (т.е. представят само имена на определени категории, например "Семейно положение").

Една типична задача за подобен тип данни е намирането на зависимости между различните променливи. Например, би било интересно да се види доколко добре доходът на човек може да бъде предсказан от стойностите на други променливи.

Разгледаната по-горе $n \times p$ матрица с данни е често само идеализация или опростяване на ситуацията, наблюдавана на практика. Например, в едно множество от медицински записи една и съща променлива (да кажем кръвно налягане) може да има множество от стойности, като всяка стойност отразява измерване, направено в различен ден. Някои пациенти могат да имат данни, представени във вид на изображение, например рентгенова снимка. Възможни са и данни във вида на текст, например диагнози и коментари на специалиста. Освен това са възможни йерархични релации между пациентите в термините на доктори, болници и географското разположение.

Очевидно е, че колко по-сложни са структурите данни, толкова по-сложни ще бъдат ИЗД моделите и алгоритмите които трябва да се използват. Макар че много от реалните множества данни не пасват точно на описания прост "плосък" (flat) формат във вид на матрица от данни, повечето от съдържащата се в тях информация може по принцип да бъде запазена и в "плоския" формат чрез подходящо дефиниране на $p$ променливи. В хода на дипломната работа оттук нататък ще подразбираме, че наблюдаваните данни съществуват във вид на $n \times p$ матрица от данни, като ще смятаме, че и $n$ и $p$ могат да бъдат много големи. В различни контексти матриците от данни се наричат с различни имена, сред които множество от данни, обучаващи данни, извадка, база от данни и т.н. В настоящата дипломна работа ще се придържаме към термина *извадка*.

## 1.3. Намаляване на обема на данните

В повечето реални случаи данните, подлежащи на анализ, имат много голям обем, което значително усложнява работата на изследователя. Първо, обработката на такива данни заема много време, което прави самият анализ практически непригоден или дори неосъществим. Второ, изследователят е ограничен в избора на "инструменти" – алгоритмите за ИЗД, тъй като не всички те еднакво добре се справят с големи обеми данни. Методите за намаляване на обема на данните целят да бъде получено такова редуцирано представяне на данните, което позволява да бъдат получени от него същите (или почти същите) аналитични резултати, както и от пълния обем данни.

Една база от данни, подлежаща на анализ, може да съдържа стотици атрибути, много от които могат да бъдат несъществени или напълно излишни за конкретната ИЗД задача. Например, ако задачата е да бъдат класифицирани пациентите на един общопрактикуващ лекар от гледна точка на това, кои от тях е най-вероятно да развият хипертония, такъв атрибут като *Телефонен_номер* на пациента е съмнително да бъде съществен, за разлика от атрибутите *Възраст* и *Тегло*. Макар че е възможно изборът на някои полезни за задачата атрибути да бъде направен от експерти в конкретна предметна област, това си остава една доста сложна и много продължителна дейност, особено когато поведението на данните не е добре изучено. Изхвърлянето на някои съществени атрибути, както и оставянето на несъществени, може да доведе до сериозни проблеми при прилагането на избрания ИЗД алгоритъм и следователно, до намаляване на качеството на извлечените закономерности. Освен това, допълнителният обем данни, предизвикан от използване на несъществени атрибути, може да доведе и до забавяне на ИЗД процеса.

Един начин за намаляване на размерността на данните е чрез премахване на несъществените атрибути. Обикновено, за целта се използват знанията на експерт в конкретна предметна област или автоматизирани методи за избор на подмножество от атрибути. Целта на избора е да бъде намерено такова минимално подмножество от атрибути, че полученото разпределение по класове на данните с използване само на тези атрибути, да е колкото се може по-близко до разпределението на данните по класове, изчислено с използване на всички оригинални атрибути. Анализирането на намалената по този начин извадка дава и още едно допълнително предимство – получените след ИЗД процеса модели съдържат по-малко количество атрибути, което ги прави по-разбираеми за потребителя.

## 1.4. Конструиране на извадки

За конструиране на извадки от големи бази от данни, които са подходящи за прилагане на ИЗД алгоритми, е важно да се вземат предвид всички описани проблеми, свързани с естеството на данните и начините за намаляване на техния обем. Самият процес на получаване на извадка може да се разглежда като процес на подготовка на „суровите" данни и получаване на две множества – множество данни и описание на множеството данни (мета-данни).

Разбиването на основната задача за подготовка на данни на подзадачи, както и съответните резултати, са представени на фигура 1.1 и са описани в таблица 1.2.

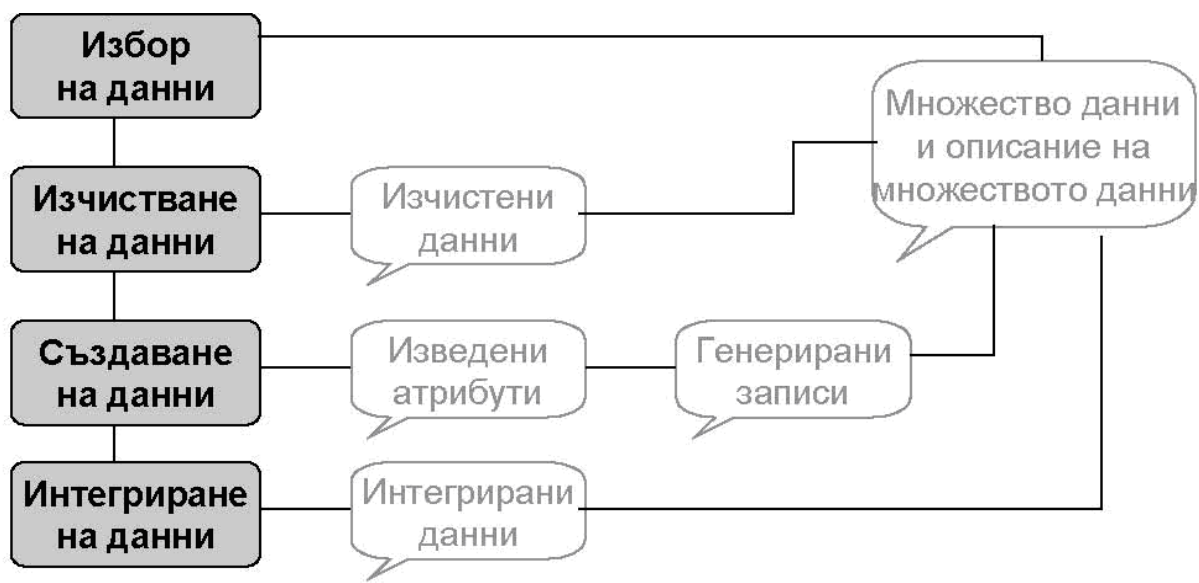

*Фигура 1.1. Етапи на подготовката на данни*

| **Избор на данни** | На тази стъпка трябва да се вземе решение, кои данни ще се използват за анализа. Критерият включва релевантността на данни за ИЗД цели, качеството на данни и техническите ограничения на обем и типове данни, които могат да бъдат обработени от наличните ИЗД алгоритми. |
|---|---|
| **Изчистване на данни** | Целта е да бъде повишено качеството на използваните данни. Това може да бъде постигнато чрез, например, избор на "чисти" подмножества от данни или чрез прилагане на съответните техники за премахването на шума и обработка на пропуснатите стойности. |
| **Създаване на данни** | Тази задача включва конструиране на нови данни чрез създаване на нови (изведени от първоначалните) атрибути и/или различни трансформации на данни от типа на нормализиране, дискретизиране и т.н. |
| **Интегриране на данни** | На този етап информацията се комбинира от различни източници (таблици) с цел да бъдат създадени нови записи или стойности. |

*Таблица 1.2. Етапи на подготовката на данни*

## 1.5. Клъстеризация на извадки

Задачата за клъстеризация (класификация) има за цел разбиването на данните на интересни и смислени подгрупи (клъстери или класове). Всички членове на една подгрупа поделят общи характеристики, например, при анализа на медицински пациенти съвкупността от поставени диагнози може да бъде разбита на отделни клъстери в зависимост от съдържащи се в него пациенти. Алгоритмите за клъстеризация събират обектите в подгрупи (клъстери), базирайки се на принципа за максимизиране на вътре-клъстерното сходство и минимизиране на между-клъстерното сходство. Клъстеризацията може да се използва и за създаване на таксономии, т.е. организацията на обекти в йерархия от клъстери, които групират заедно сходните обекти.

Клъстеризацията може да се извършва ръчно или (полу-)автоматично. Изследователят може да направи предположение за броя на сегментите, използвайки предварителните основни знания за проблемната област или базирайки се на резултати от описанието и обобщението на данните, а ИЗД алгоритъм да извърши сегментирането на данните на указания брой сегменти. Съществуват и техники за автоматична клъстеризация, които сами могат да намерят по-рано неизвестни и скрити структури в данните без човешка намеса.

Клъстеризацията може да бъде напълно самостоятелна цел на ИЗД изследване, т.е. определяне на сегментите да бъде основната цел на ИЗД процеса. Обаче, доста често тя е само една стъпка при решаване на други ИЗД задачи. В тези случаи целта на клъстеризацията е да осигури размер на данните, удобен за други ИЗД цели, или да намери еднородни подмножества от данните, които след това могат много по-лесно да бъдат анализирани. Обикновено в големи бази различни зависимости между данните влияят едни върху други и затрудняват намирането на интересни закономерности. В този случай подходящото клъстеризиране на данните значително улеснява основната задача.

В настоящата дипломна работа няма да се спираме на различни методи за клъстеризация. Ще считаме, че извадките са предварително клъстеризирани (класифицирани), като ще използваме принадлежността на обектите към техния клъстер (клас) за по-информативно визуализиране на данните.

## 1.6. Визуализация на извадки

Задачата за визуализация на извадки цели намирането на стегнато представяне на основните характеристики на данните, обикновено в елементарна, обобщена форма. Това позволява на потребителя да получи представа за структурата на данните.

В повечето случаи визуализацията на данните е само една от подзадачите, изпълнявана на ранните стадии от едно ИЗД изследване. В началото на изследването потребителят често не знае нито точната цел на анализа, нито естеството на данните. Началният изследователски анализ на данните може да помогне да бъде разбрана природата на данните, да бъдат построени първоначалните хипотези за скритата в данните информация. Например, разпределението на диагнозите на пациентите спрямо мястото на живеене може да даде насока за това, кои групи пациенти са застрашени повече от развиване на дадено заболяване и това да доведе до взимане на нови превантивни медицински мерки. По тази причина разгледаният в настоящата дипломна работа метод за визуализация е част от ИЗД етапа на разбиране на данните.

Работещите в медицинската сфера – общопрактикуващи лекари и специалисти – са свикнали с графичното представяне на информацията, защото за медицинската информация това е един по-разбираем и полезен начин за представяне на данните. Ето защо в настоящата дипломна работа се предлага графична визуализация на извадките. Като удачен метод за това е избран алгоритъмът FastMap [Faloutsos & Lin, 1995], който е разработен за целите на бързото търсене в мултимедийни бази от данни, както и за визуализацията на многомерни данни.

Алгоритъмът FastMap решава задачата за проектиране на *N* обекта, за които е известна *N x N* матрицата на взаимните разстояния, в *N* точки в *k*-мерно пространство по начин, запазващ (до голяма степен) съответствията в разстоянията между обектите. Решаването на задача за проектиране на *N n*-мерни обекта в *N k*-мерни обекта, където $k \leq n$, е частен случай на тази по-обща задача. В настоящата дипломна работа ще използваме проектиране с *k = 2*, т.е. проектиране в двумерно пространство.

Вместо матрица на взаимните разстояния може да бъде използвана някаква мярка за разстояние между два обекта, дефинирана като функция D(A, B) := разстоянието между обектите A и B. В оригиналната версия на алгоритъма се използва Евклидово разстояние, но за нашите нужди ще дефинираме по-различни мерки за разстояние, които са съобразени с типовете на атрибутите от медицинските извадки.

## *1.7. Цел на дипломната работа*

След като представихме накратко целта и методите за ИЗД, описахме основните операции по конструиране и визуализиране на извадки, вече можем по-точно да дефинираме целта на настоящата дипломна работа:

Целта на дипломната работа е проектиране и създаване на софтуерна система, която е съставена от два отделни инструмента:

1. „Конструктор на извадки", с чиято помощ един лекар да може лесно да конструира извадки от големи медицински бази от данни по дефиниран от него критерий. Тези извадки представляват многомерни, класифицирани медицински данни с подходящо описание (мета-данни). Конструирането трябва да притежава интерактивни свойства, които да позволяват намаляване обема да данните чрез задаване на критерий за избор на подходящи обекти и техните атрибути.

2. „Визуализатор на извадки", с чиято помощ така конструираните извадки да могат да се визуализират в двумерно пространство чрез намаляване размерността на данните по FastMap алгоритъм. Визуализацията трябва да притежава интерактивни свойства, които да позволяват лесното разглеждане на данните, стоящи зад отделните елементи от визуализацията.

Целта на дипломната работа е постигната, като са проектирани и разработени две програми: *Konstruktor* и *Vizualizator*, които изпълняват функционалните изисквания към двата инструмента.

# 2. Проектиране на Конструктора

Конструкторът на извадки трябва да позволява на всеки лекар лесно да конструира извадки от големи медицински бази от данни. При това лекарят трябва да може сам да дефинира критерии, на които иска да отговарят обектите в получената извадка. В тази глава ще разгледаме структурата на големите бази от данни и ще проектираме начин, по който интерактивно да могат да се конструират извадки от тях.

## *2.1. Организация на данни и бази от данни*

Една от характеристиките, която отличава извличането на закономерности от данни от други типове аналитични задачи, е количеството на данните. В много от ИЗД приложенията матрицата от данни съдържа милиони редове и хиляди колони. По тази причина въпросите за ефективността на алгоритмите за обработка на данните са от много съществено значение. Един алгоритъм, чието време за изпълнение нараства експоненциално с нарастването на броя $n$ на редовете, може да бъде неизползваем за всички практически случаи, освен за много малки множества от данни.

Във всеки ИЗД проект е полезно да се прави разграничение между две фази. По време на първата се подготвят данните за ИЗД алгоритъма, а по време на втората се изпълнява самият алгоритъм. Първата фаза изглежда тривиална, обаче често точно тя се оказва "тясното място" в проекта. Например, при анализа на данни често е необходимо избраният ИЗД алгоритъм да бъде прилаган многократно към различни подмножества от данните. Това означава, че ние трябва да сме в състояние бързо да намираме членовете на всяко желано подмножество, както и да извличаме от базата и да зареждаме това подмножество в основната памет.

Целта на Конструктора на извадки е именно такъв инструмент, който да дава бърз и лесен начин за извличане на различни подмножества от данните.

## *2.2. Релационни бази от данни*

Повечето съвременни медицински информационни системи използват релационни бази от данни (RDB) за съхранение на огромния обем от медицинска информация. Системите за управление на релационни бази от данни (RDBMS) предоставят един унифициран механизъм за бърз достъп към избрани части от данните в базата.

Релационният модел на данни се базира върху идеята за представяне на данните в таблична форма. Схемата на една таблица се състои от името на таблицата и множество от именувани колони. Имената на колоните се наричат променливи или атрибути. Действителната таблица (един екземпляр на схемата), наричана още релация, представлява едно именувано множество от редове. Всеки елемент на една таблица съдържа в колонката за атрибута A някаква стойност от дефиниционната област

Dom(A) на колоната A. Един атрибут може да бъде от произволен тип – цели или реални числа, низове, дати и т.н. Подредбата на редовете и колоните в таблицата не е от значение.

Данните във всяка клетка на една таблица обикновено са атомарни стойности, т.е. не е прието използването на списък от стойности за една клетка. Това означава, че ако искаме да представим информацията за хора, тяхната възраст и телефонни номера, ние не можем да пазим в един атрибут няколко телефонни номера на един и същ човек. Ограниченият по този начин модел има така наречената *първа нормална форма*. При проектиране на Конструктора от настоящата дипломна работа ще считаме, че базата от данни, от която ще се правят извадки, отговаря на условията за първа нормална форма.

Дори в относително малки организации, използваните релационни бази от данни могат да имат стотици таблици и хиляди атрибути. По тази причина, управлението на схемата на базата от данни може да се окаже една доста сложна задача. Понякога се твърди, че за целите на анализа на данни е достатъчно всички тези таблици да бъдат комбинирани в една масивна таблица с наблюдения или "универсална таблица", като по този начин ИЗД анализаторът не трябва да се грижи, че данните се намират някъде в база от данни. Обаче, проверката на тази теза дори върху прости примери показва, че това не е осъществимо: универсалната таблица ще бъде толкова голяма, че операциите върху нея ще бъдат извънредно скъпи.

## *2.3. Език на структурирани заявки (SQL)*

За управлението на данните не е достатъчно само да можем да описваме структурата им и да ги съхраняваме с използване на тази структура. Трябва ни още възможност да извличаме данни по желание от създадената база. Ще разгледаме накратко езика на структурните заявки (SQL), който предлага удобен механизъм за извличане на данни от базата, като използва логически условия за дефиниране на критерии за търсене.

В системите за управление на бази от данни SQL е стандартът, приет от повечето от производители на подобни системи. SQL реализира едно надмножество от релационната алгебра, като по този начин осигурява една удобна и компактна нотация за боравене с множества от данни. В настоящия раздел ще разгледаме само базовата структура на SQL програмите.

Базовата конструкция на SQL е израз или заявка от типа на "select-from-where" (избери-от-където), която има обикновено следната най-проста форма:

**select** $A_1, A_2, \ldots, A_p$
**from** $r_1, r_2, \ldots, r_k$
**where** списък от условия

Тук всяка $r_i$ е таблица, а всеки $A_j$ е атрибут. Интуитивното значение е, че за всеки възможен избор на редове $t_1, \ldots, t_k$ от таблиците $r_1, r_2, \ldots, r_k$ ние проверяваме, дали условията са верни. Ако да, то редът, състоящ се от стойностите на атрибутите $A_j$, се връща в резултата.

Вторият ред на заявката - клаузата **from** - описва таблиците, към които тази SQL конструкция трябва да бъде приложена. Третият ред, съдържащ клаузата **where**, описва условията, на които редовете на тези таблици трябва да отговарят, за да бъдат включени в резултата. Първият ред - клаузата **select** - определя кои атрибути от указаните таблици трябва да фигурират в резултата. Тя съответства на операцията *проектиране* (π) от релационната алгебра (не на операцията за избор - σ). Клаузата **where** се използва за представяне на условията за избор, участващи в операциите *избор* и *съединение*. За операцията *избор* условията за избор са представени като списък от условия в клаузата **where**, разделени от такива ключови думи като **and**, **or** или **not**.

Например, избор на всички пациенти от таблицата с име "Пациенти", които са на възраст над 20 години, може да бъда направен чрез заявката:

**select** пациент
**from** Пациенти
**where** възраст > 20

Ако някои таблици от клаузата **from** имат общи атрибути, имената на атрибутите трябва да имат префикс, съставен от точка и име на таблицата, когато те се появяват в клаузата **select** или **where**.

Например, намирането на всички прегледи от таблицата с името "Прегледи", които са били извършени на пациенти над 20 години, може да се направи със следната заявка:

**select** Пациенти.пациент, Прегледи.преглед, Пациенти.възраст
**from** Пациенти, Прегледи
**where** (Прегледи.пациент = Пациенти.пациент)
**and** (Пациенти.възраст > 20)

Тук изреждането „Пациенти, Прегледи" всъщност представлява вътрешно свързване (**inner join**) на двете таблици по подразбиращите се атрибути. Например по полето ЕГН на пациента, което трябва да е еднакво за съответстващите записи в базата на един пациент и всички негови прегледи. Особеност на вътрешното свързване е, че в резултата ще присъстват само тези пациенти, които са имали поне един преглед. Ако обаче искаме да получим всички пациенти над 20 години, без значение дали са имали преглед или не, тогава трябва да използваме външно свързване (**outer join**):

**select** Пациенти.пациент, Прегледи.преглед, Пациенти.възраст
**from** Пациенти
**left outer join** Прегледи **on** (Прегледи.пациент = Пациенти.пациент)
**where** (Пациенти.възраст > 20)

Ако всички атрибути от участващите в заявката таблици трябва да фигурират в резултата, то списъкът от атрибути в клаузата **select** може да бъде заместен със звездичка (*).

Тъй като по подразбиране записите в релационната база от данни нямат фиксирана наредба, затова се използва клауза **order by** за указване на начина на сортиране на резултата. Например, ако искаме да получим всички пациенти на възраст между 30 и 50 години, като ги подредим по намаляване на възрастта, ще използваме следната SQL заявка:

```
select пациент
from Пациенти
where (възраст >= 30) and (възраст <= 50)
order by възраст desc
```

Тук модификаторът **desc** указва да се използва сортиране в низходяща последователност.

### 2.4. Изпълнение и оптимизация на заявки

Една заявка може да бъде изпълнена по няколко различни начина. Тривиалният метод за изпълнение е да опитаме всички възможни двойки от редове от цитираните таблици, за да проверим, дали те отговарят на логическия израз в **where** клаузата.

Един по-ефективен метод е да се използват допълнителни индекси (например B-дървета и хеш таблици), чрез които да се изключат бързо голяма част от записите и да се прегледат само малка част от тях дали наистина отговарят на критериите. Това може да се направи например за условието „**where** възраст > 20", стига да има дефиниран индекс по атрибута *възраст* в базата.

Оптимизацията на заявки е задача за намиране на възможно най-добър метод за изпълнение на една конкретна заявка. Обикновено, оптимизаторите на заявки транслират SQL-заявката в едно дърво на изразите, в което листата представляват таблици, а междинните възли – операции върху наследниците на възлите. Алгебрични равенства между операциите могат да се използват за трансформация на дървото в някоя еквивалентна форма, която се изпълнява по-бързо. След намиране на подходящо дърво на изразите се избира метод за изпълнение на всяка операция. Например, операцията съединение (**join**) може да бъде изпълнена по няколко различни начини: чрез вложени цикли, чрез сортиране или чрез използване на индекси. Ефективността на всеки метод зависи от размера на таблиците и разпределението на стойностите в тези таблици. Ето защо оптимизаторите на заявки пазят информация за разпределението на стойностите на отделните атрибути, за да намират добри методи за изпълнение на заявки. Теоретично, намирането на най-добрата стратегия за изпълнение за една конкретна заявка е NP-пълна задача, така че намирането на най-добрия метод не е осъществимо. Обаче, добрите оптимизатори на заявки могат да бъдат учудващо ефективни.

Системите за управление на бази от данни се стремят да предоставят добро изпълнение за един широк диапазон от заявки. Макар че за една единствена заявка е възможно да бъде написана специална програма, която изчислява резултата по-ефективно от начина, по който го прави системата за управление на бази от данни, силата на подобни системи е в това, че те осигуряват бързо изпълнение за повечето заявки. Това е изключително полезно за нуждите на Конструктора, тъй като заявките за ИЗД обикновено не са известни предварително. Точно по тази причина вместо да реализираме частичен механизъм за изключително бързо изпълнение на няколко конкретни заявки, предпочитаме да използваме за нуждите на Конструктора езика SQL за достатъчно бързо изпълнение на много по-широк клас от заявки.

## 2.5. Проектиране на потребителския интерфейс

Дефинирането на критериите, на които трябва да отговаря извадката, е най-предизвикателната част от Конструктора на извадки. По принцип, потребителският интерфейс трябва да скрива сложността на лежащия в основата синтаксис на SQL. В случая с **where** клаузата това е много трудно за постигане, защото целта е да се използва пълния потенциал на сложния логически формализъм на SQL.

Един възможен подход за реализиране на конструктор на SQL изрази е да се започне с обикновен текстов редактор, към който да се добавят допълнителни функции, улесняващи писането на логическите изрази в **where** клаузата. Това са функции като синтактично оцветяване (syntax highlighting), автоматично дописване (auto-completion) и проверка на изразите за правилност. Такъв подход (виж фигура 2.1) е използван в редактора SQL Editor, който е част от IB Expert системата за администриране на релационни бази от данни [IB Expert, 2006].

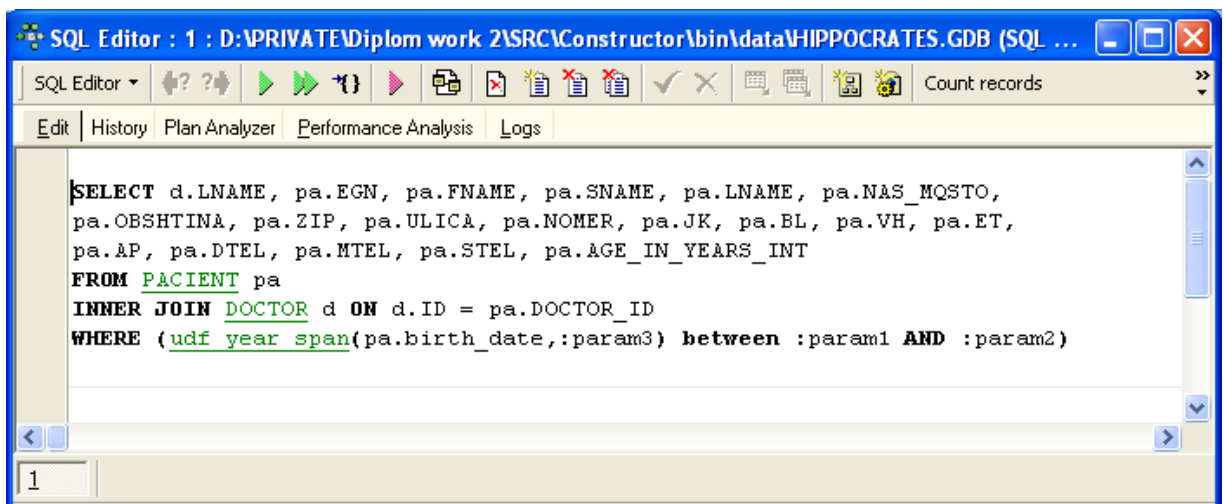

*Фигура 2.1. Пример за надстроен текстов SQL конструктор на заявки*

Очевидно, такъв подход може да се използва само за най-напредналите потребители, които знаят отлично синтаксиса на езика SQL, знаят „наизуст" имената на таблиците и атрибутите им и се справят с директното записване на логическите изрази в текстов вид. Този подход обаче е напълно безполезен за по-неопитни потребители, каквито очакваме да бъдат потребителите на Конструктора на извадки.

Ясно е, че трябва да се проектира такъв процес на конструиране, който да позволи колкото се може по-пълно и едновременно с това достатъчно лесно използване на SQL езика. Затова избираме като интерфейс да използваме падащи списъци, избор от изброени варианти, таблично представяне и поредово редактиране на критериите. Пример за подобен интерфейс е даден на фигура 2.2. В допълнение, предвиждаме възможност за работа с различно подмножество (т.е. изразителна сила) на езика SQL, в зависимост от това колко е напреднал потребителя.

*Фигура 2.2. Пример за интерфейс с използване на падащи списъци и таблично представяне на условията в SQL конструктор на заявки*

## 2.6. Проектиране на процеса на конструиране

Тъй като се предполага, че потребителите на Конструктора няма да са специалисти в областта релационните бази, основната цел на инструмента е да се абстрахира максимално от синтаксиса на SQL и едновременно с това да гарантира синтактичната коректност на конструираните SQL заявки.

Предлагаме два режима на работа, съобразени с опитността на потребителя:

- Първият, *режим за начинаещи*, позволява конструиране само на по-прости извадки. На потребителя се предоставя възможност за избор на данни, условия за удовлетворяване, колони в резултата и експорт на данните. Този подход е подходящ за начинаещи потребители.

- Вторият, *режим за напреднали*, предоставя възможност за контрол над свързването на логическите изрази с оператори, групиране на резултата от извадката, както и повече възможности за експорт на извадката в различни файлови формати. Това прави работата на по-опитните потребители по-ефективна.

За улеснение на потребителя разделяме процеса на конструиране на една SQL заявка на няколко стъпки, съобразени със структурата на езика за заявки. На всяка стъпка потребителят може да извършва само такива действия, които се отнасят до определена подчаст на конструираната SQL заявка. Кратко описание на проектираните стъпки е дадено в таблица 2.1. Със звезда са означени тези стъпки, които са налични само в *режима за напреднали*.

| Стъпка | Наименование | Описание |
| --- | --- | --- |
| 1 | Данни за извадката | Отговаря за **from** клаузата на заявката и за **join** клаузата за свързване на допълнителни таблици |
| 2 | Условия за удовлетворяване | Отговаря за **where** клаузата на заявката и за логическите оператори в нея (**and, or, not**) |
| 3 | Колони в резултата | Отговаря за **select** клаузата на заявката в частта избор на атрибути (колони) в резултата |
| 4 | Сортиране на редовете | Отговаря за **order by** клаузата на заявката и избор на възходящ/низходящ ред |
| 5* | Фина настройка | Допълнителни настройки |
| 6 | Резултат от извадката | Визуализира резултата в табличен вид |
| 7* | Групиран резултат | Предлага възможност за групиране на редовете в резултата по един или повече атрибути |
| 8* | Експорт на резултата | Предоставя експорт на (евентуално групирания) резултат в няколко от най-популярните формати за обмен на данни. |

*Таблица 2.1. Кратко описание на проектираните стъпки*

По наши наблюдения, основните проблеми, които възникват по време на конструирането на една извадка, са свързани с използването на правилните имена на таблиците и атрибутите, както и със задаване на логическите условия за удовлетворяване, съобразени с типовете на атрибутите.

Конструкторът се справя с тези проблеми по три начина:

- Първо, потребителят е ограничен в избора си – той може да избира само атрибути от наличните в избраните таблици и не може да пише произволни наименования на ръка.

- Второ, разрешено е добавянето само на коректни логически условия върху избраните атрибути, съобразени с типовете на съответните колони.

- Трето, логическите връзки между отделните условия за удовлетворяване се генерират автоматично от Конструктора на базата на евристични правила и проверки, които в повечето случаи съвпадат с истинското намерение на потребителя. Това „умно" поведение на Конструктора може да бъде управлявано по-прецизно в режима за напреднали.

В резултат, създадените SQL заявки винаги са синтактично коректни спрямо схемата на релационната база, от която се извличат данните.

### 2.2.1. Стъпка 1 - „Данни за извадката"

В тази стъпка се указват таблиците, от които ще се прави извадката. Това включва дефиниране както на **from** клаузата на заявката, така и евентуални **join** клаузи за свързване на допълнителни таблици.

Тъй като основната цел на Конструктора е да се абстрахира максимално от синтаксиса на SQL, затова вместо да се показват имена на таблици и ключове за свързването им, в Конструктора се показват списък с предварително дефинирани варианти за избор. Тези варианти представляват готови множества от **from** и **join** клаузи, дефинирани предварително от медицински експерт с помощта на администратор на базата от данни. На фигура 2.3 е показана как изглежда тази стъпка в Конструктора.

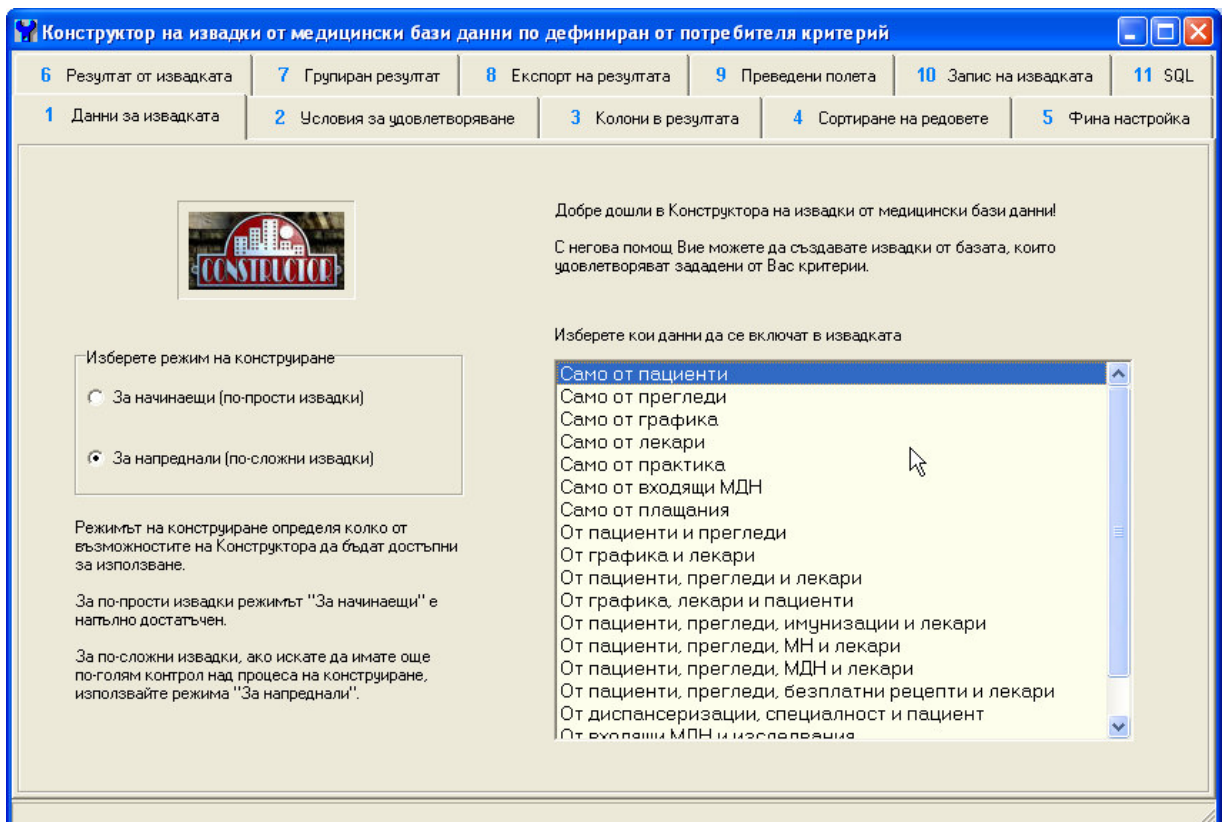

*Фигура 2.3. Стъпка 1 - „Данни за извадката"*

Ако потребителят избере „Само от пациенти" от предефинираните варианти, то в този момент ще се генерира приблизително следната първоначална SQL заявка:

```
select *
from Пациенти
```

Аналогично, ако потребителят избере „От пациенти и прегледи", то SQL заявката ще изглежда приблизително така:

```
select *
from Пациенти
inner join Прегледи on (Прегледи.пациент = Пациенти.пациент)
```

Спестяването на детайлите относно ключовите полета и свързването на таблиците е огромно облекчение за крайните потребители, тъй като познаването на тези подробности е по силите само на опитен администратор на бази.

### 2.2.2. Стъпка 2 - „Условия за удовлетворяване"

Това е може би най-интересната стъпка от конструирането, защото тя отговаря за задаване на критериите, на които трябва да отговарят обектите от извадката. Критериите се представят чрез логическите условия в **where** клаузата на заявката, както и свързващите ги логически оператори - **and, or** и **not**. На фигура 2.4 е показан общ изглед от тази стъпка в Конструктора.

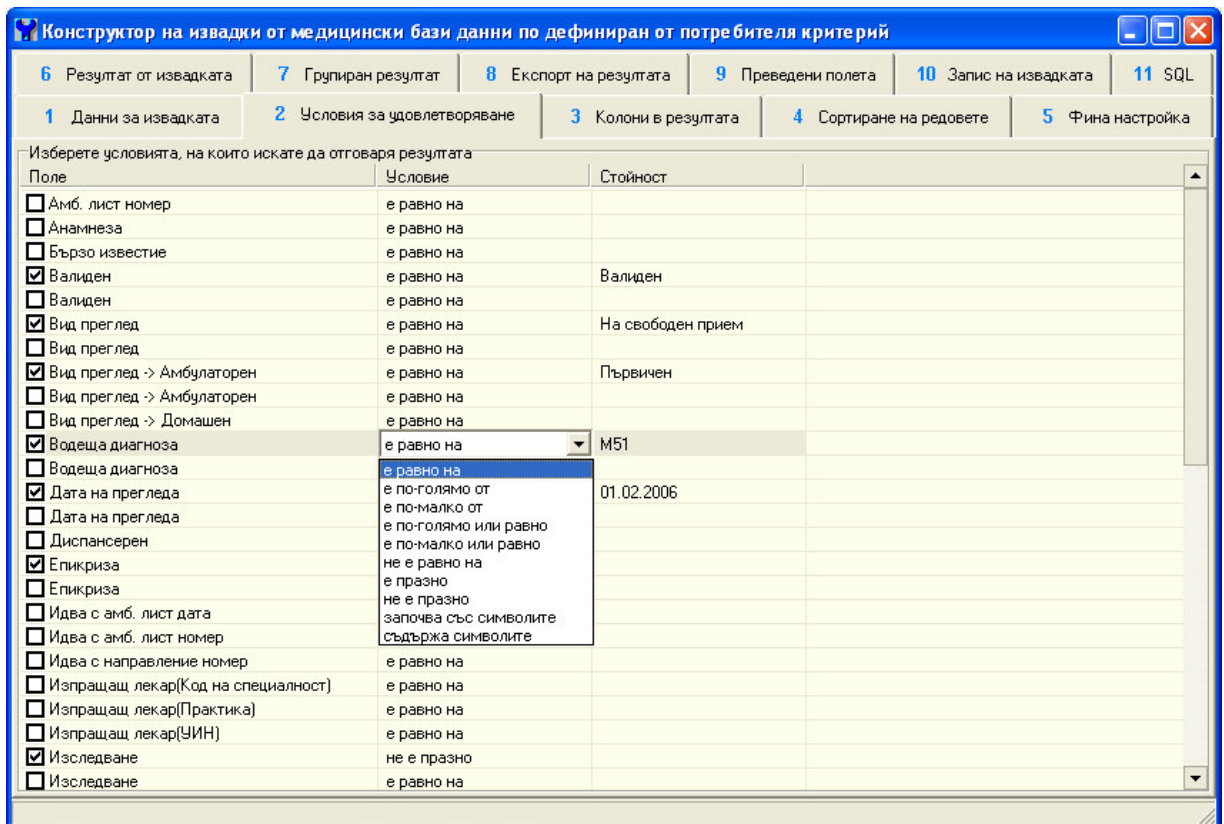

*Фигура 2.4. Стъпка 2 - „Условия за удовлетворяване"*

За максимално улеснение на потребителите сме проектирали следните ключови функционални възможности за тази стъпка:

- Първо, на потребителят се предоставя изчерпателен списък на наличните атрибути, върху които могат да се налагат логически условия. Така потребителят е ограничен в избора си и не може да пише произволни наименования на ръка (и съответно да допуска грешки). Потребителят просто трябва да маркира тези атрибути, върху които иска да наложи някакви логически ограничения, например по този начин:

- Второ, за всеки атрибут се допуска задаване само на такива логически операции, които са допустими за съответния тип на атрибута. Нещо повече, въвеждането на допълнителни операнди (най-често константи) преминава през допълнителна проверка, която гарантира съвместимостта на използваните типове данни. Например, за атрибут от числов тип се допускат операции като по-голямо, по-малко, равно, различно и съответно се допуска въвеждане на операнди от числов тип. В таблица 2.2 са описани типовете атрибути и съответните операции и типовете операнди, които се допускат за тях.

*Таблица 2.2. Съобразени с типовете на атрибута допустими операции и операнди*

| Тип на избрания атрибут | Допустими операции | Допустими операнди |
|---|---|---|
| Булев тип (True/False) | е равно на / не е равно на / е празно / не е празно | Маркирано / Немаркирано |
| Числов тип | е равно на / е по-голямо от / е по-малко от / е по-голямо или равно / е по-малко или равно / не е равно на / е празно / не е празно | Числов тип, съвпадащ с числовия тип на атрибута (целочислен или с плаваща запетая) |

| Тип на избрания атрибут | Допустими операции | Допустими операнди |
|---|---|---|
| Символен тип | 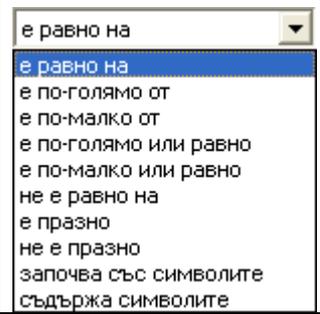 е равно на / е по-голямо от / е по-малко от / е по-голямо или равно / е по-малко или равно / не е равно на / е празно / не е празно / започва със символите / съдържа символите | Произволен символен тип |
| Календарен тип (дата) | 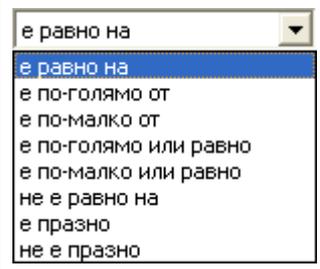 е равно на / е по-голямо от / е по-малко от / е по-голямо или равно / е по-малко или равно / не е равно на / е празно / не е празно | Дата, избрана от календар, така че да е валидна. 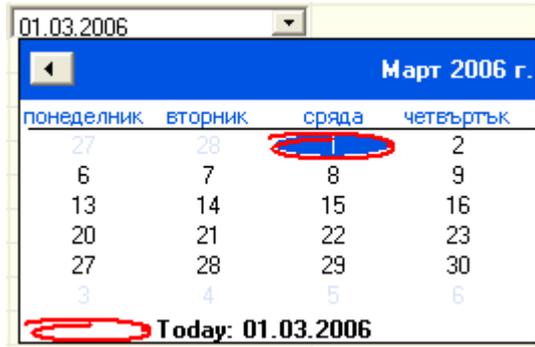 |

*Таблица 2.2. Съобразени с типовете на атрибута допустими операции и операнди*

Операцията „е празно" съответства на SQL операцията (IS NULL), което е съвсем различно от условието за равенство с празен низ (= "").

- Трето, всеки атрибут, който е включен в логически израз автоматично получава свое ново копие (клонинг), който е свободен за свързване в нови операции. По този начин могат да се конструират сложни логически условия върху един и същи атрибут. Например, задаване на възрастов диапазон от 30 до 50 години включва две условия върху атрибута възраст: (възраст >= 30) и (възраст <= 50). След задаване на всяко от тези две условия автоматично Конструкторът създава ново копие на атрибута *Възраст*, за да може да се използва за ново условие. В резултат се получават три екземпляра на атрибута, две от които са свързани с конкретно условие, а третият е свободен за използване:

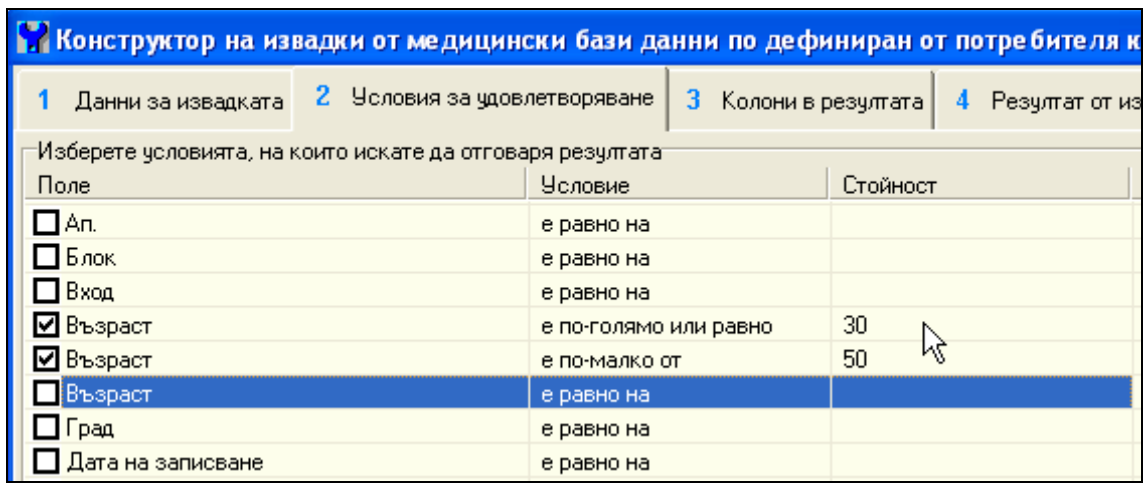

- Четвърто, потребителят се освобождава от задължението явно да задава логическите връзки между отделните логически условия. Специално проектирания механизъм за „умно свързване" на логическите изрази се грижи да постави правилната операция на правилното място – **AND или OR**. Най-важните евристични правила, които се използват за реализиране на „умното свързване", са изброени в таблица 2.3.

*Таблица 2.3. Евристични правила за „умно свързване" на логически условия"*

| Първо условие | Второ условие | Операнди | „Умно свързване" |
|---|---|---|---|
| Атрибут = A | Атрибут = B | A <> B | **OR** |
| Атрибут > A или Атрибут >= A | Атрибут < B или Атрибут <= B | A <= B | **AND** 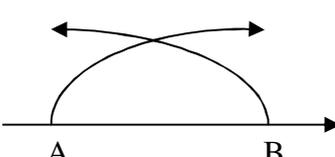 |
| Атрибут IS NULL или Атрибут IS NOT NULL | Атрибут = A | - | **OR** |
| Атрибут <> A | Атрибут <> B | - | **AND** |
| Атрибут > A или | Атрибут < B или | A > B | **OR** |

| Атрибут >= A | Атрибут <= B | | 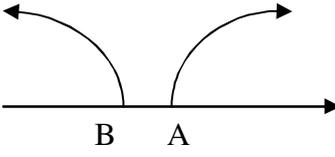 B  A |
| --- | --- | --- | --- |
| Атрибут STARTING A | Атрибут STARTING B | A <> B | **OR** |
| Атрибут < A (аналогично Атрибут > A) | Атрибут < B (аналогично Атрибут > B) | - | **AND** 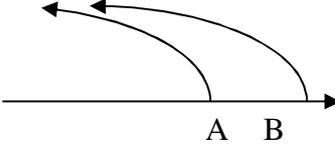 A  B |
| Атрибут LIKE A | Атрибут LIKE B | A <> B | **OR** |

Следвайки тези правила, Конструкторът съединява отделните условия, зададени от потребителя, като се получава един единствен логически израз, който се превежда на SQL езика в **where** клаузата. Например, ако потребителят зададе две условия върху атрибута възраст: (възраст >= 30) и (възраст <= 50), то на базата на тези евристични правила Конструкторът ще използва **AND** логическа връзка, за да ги свърже (вместо например **OR**, което в този случай би превърнало условието в константата **TRUE** и би върнало всички пациенти от базата).

Друг пример: потребителят е задал условие за диагнозата на пациента: (диагноза = „O12") и (диагноза = „E18"). Този път на базата на евристичните правила Конструкторът ще използва **OR** логическа връзка (вместо например **AND**, което в този случай би превърнало условието в константата **FALSE** и не би върнало нито един пациент като резултат).

### 2.2.3. Стъпка 3 - „Колони в резултата"

Дотук конструираните SQL заявки неизменно избираха абсолютно всички налични атрибути на обектите, защото започваха с клаузата:

```
select *
from ...
```

Стъпката „Колони в резултата" е втората стъпка след „Условия за удовлетворяване", която има пряко отношение към намаляване размерността на данните, още преди те да са подложени на обработка от FastMap алгоритъма.

Потребителят отново е улеснен от предварително подготвен списък със смислени наименования на атрибутите на избраните обекти, от които той може да избере интересуващото го подмножество от необходими атрибути. Интересното в случая е, че потребителят не само задава подмножество от атрибути, но задава и тяхната конкретна подредба, която се използва в групирания резултат и при експортирането на данните. Така потребителят има по-добър контрол над окончателния вид на извадката. На фигура 2.5 е показана как изглежда тази стъпка в Конструктора.

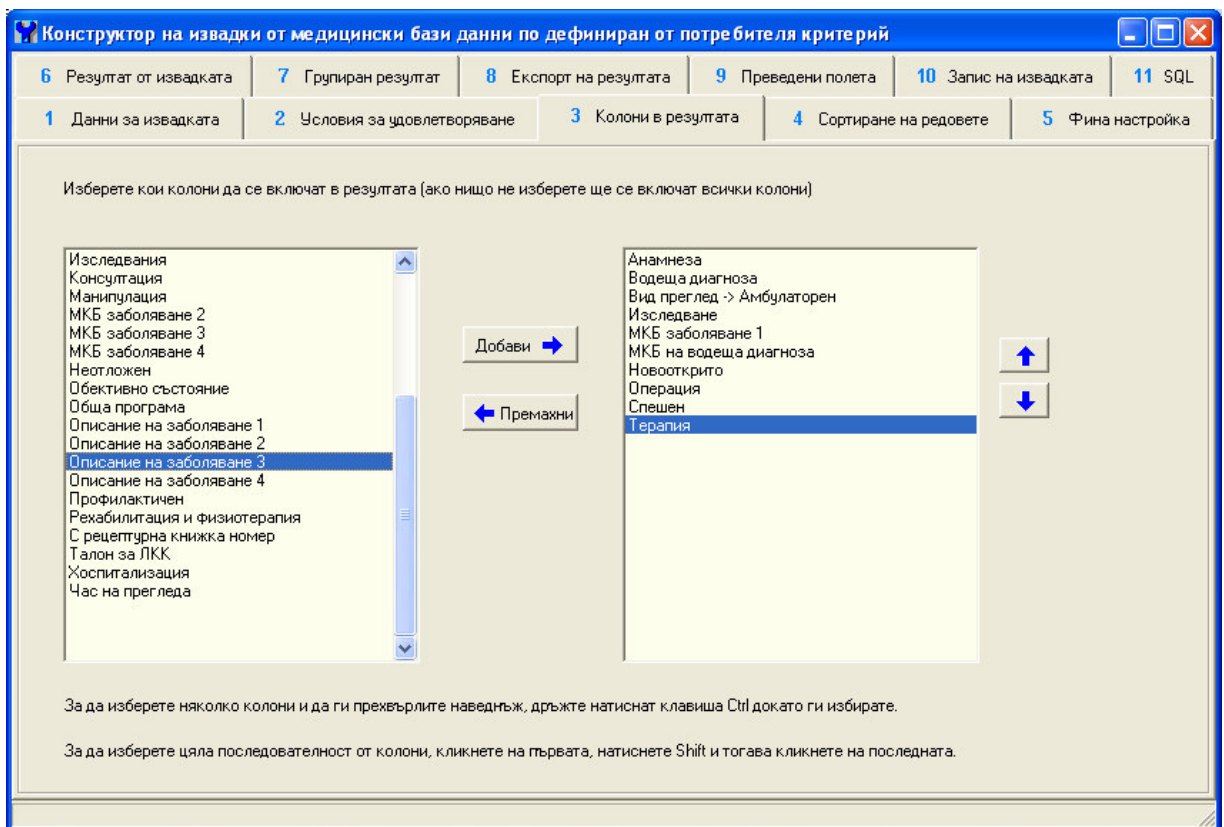

*Фигура 2.5. Стъпка 3 - „Колони в резултата"*

Съответно генерираният текст в SQL заявката изглежда така:

**select** ANAMN, MAIN_DIAG_OPIS, AMB_PR, IS_ANALYSIS, PR_ZAB1_MKB, MAIN_DIAG_MKB, IS_NEW, IS_OPERATION, IS_EMERGENCY, TERAPY
**from** PREGLED
...

## 2.2.4. Стъпка 4 - „Сортиране на редовете"

Тази стъпка няма пряко отношение към ИЗД, тъй като за Визуализацията (а и въобще за ИЗД алгоритмите) последователността на обектите не трябва да има значение. Но за по-голямо удобство за разглеждане на данните е предвидена тази възможност, те да бъдат подредени по зададени от потребителя атрибути. За всеки атрибут, по който ще се прави сортиране, се задава приоритет и посока на сортиране (възходящо/низходящо). На фигура 2.6 е показана как изглежда тази стъпка в Конструктора.

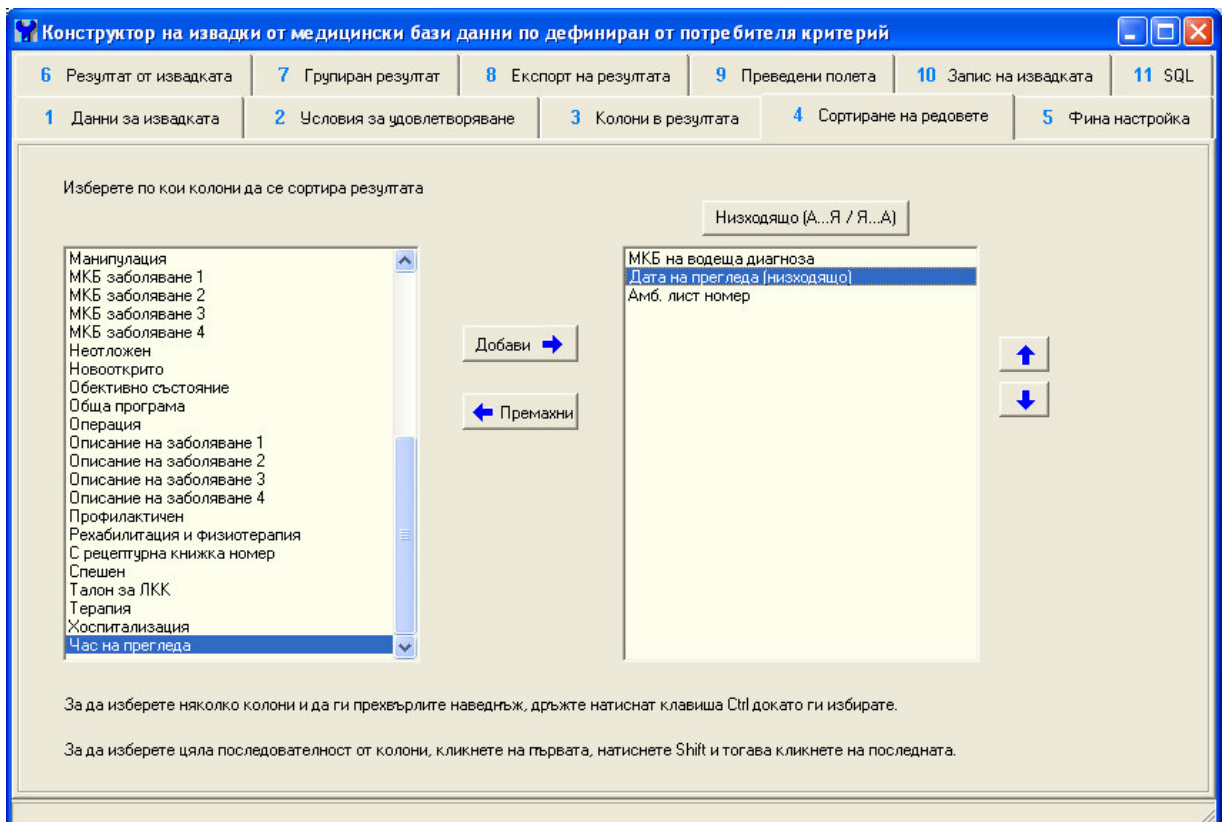

*Фигура 2.6. Стъпка 4 - „Сортиране на редовете"*

Съответно генерираният текст в SQL заявката изглежда така:

```
select ...
from PREGLED
order by  MAIN_DIAG_MKB ASC, START_DATE DESC, AMB_LISTN ASC
...
```

### 2.2.5. Стъпка 5 - „Фина настройка"

Тази стъпка (достъпна само в режима за напреднали) предоставя по-фин контрол над изпълнението на конструираните заявки. Ето някои от настройките, които се предоставят на потребителя в тази стъпка:

- Възможност за елиминиране на повторенията от извадката, т.е. включване само на уникалните редове в извадката. По този начин е възможно да се елиминира чувствителността на ИЗД алгоритмите за анализ от повторение на обекти в извадката.
- Възможност за контролиране на механизма „умно свързване" и евентуално ръчно указване на логическите оператори.
- Възможност за контрол над автоматичното именуване на колоните в резултата.

На фигура 2.7 е показана как изглежда тази стъпка в Конструктора.

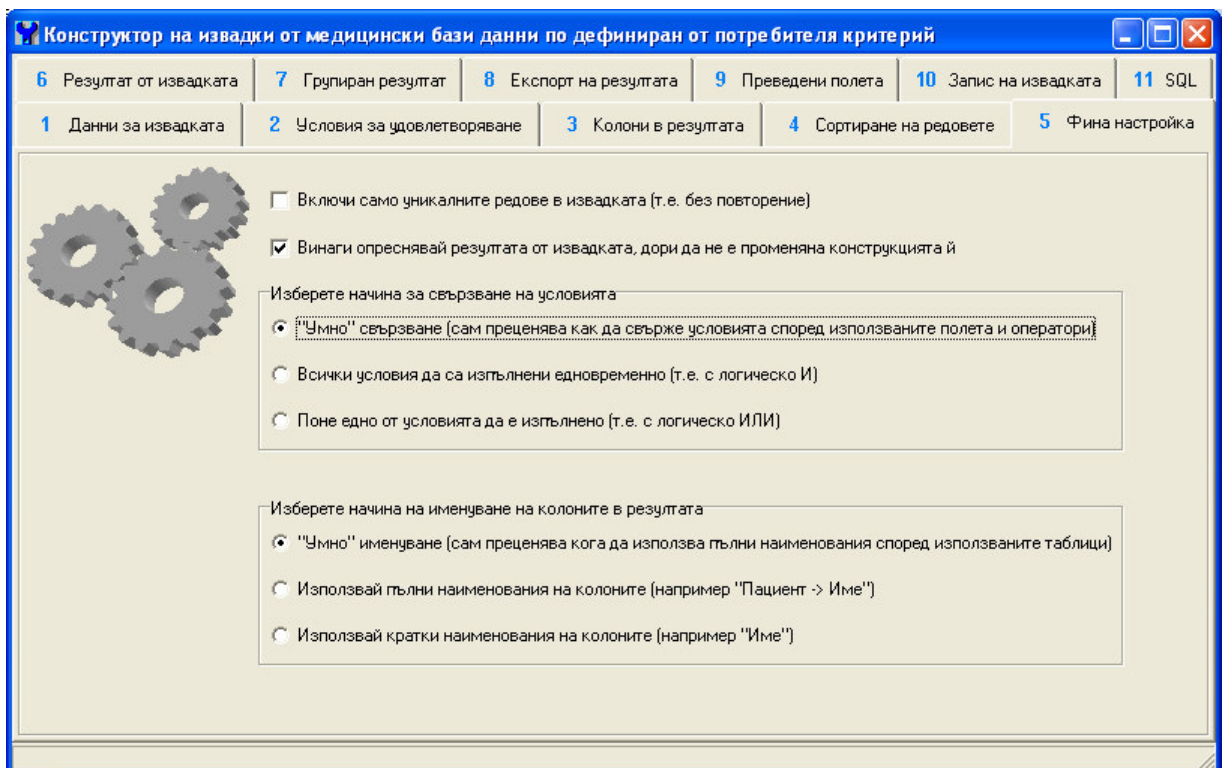

*Фигура 2.7. Стъпка 5 - „Фина настройка"*

### 2.2.6. Стъпка 6 - „Резултат от извадката"

На тази стъпка се визуализира резултатът от изпълнението на конструираната SQL заявка в табличен вид. Трябва да се има предвид, че изпълнението на (дори сравнително прости) заявки над големи бази от данни може да отнеме доста време и същевременно да изисква заделяне на доста динамична работна памет. Ето защо основното предимство на тази стъпка е, че се прави частично зареждане (partial load) на резултата. Това означава, че вместо да се извличат от базата всички записи от резултата, се извличат само толкова записи, колкото са достатъчни, за да се покаже видимата част от таблицата. След първоначалното частично зареждане потребителят има възможност да изличи още записи като използва вертикалния плъзгач (vertical scroller).

Идеята на тази стъпка е, преди да се извлече абсолютно целият резултат от базата, потребителят да има възможност да огледа получаваните обекти дали наистина са тези, които той очаква да получи. Тази стъпка спестява доста време (и памет на клиентския компютър), тъй като внимателното настройване на условията обикновено отнема поне 2-3 опита до достигане на желания резултат. На фигура 2.8 е показана как изглежда тази стъпка в Конструктора.

*Фигура 2.8. Стъпка 6 - „Резултат от извадката"*

### 2.2.7. Стъпка 7 - „Групиран резултат"

Тази стъпка започва с любезното съобщение „Моля, изчакайте..." и това не е случайно.

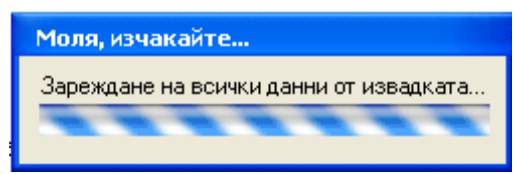

В тази стъпка се извлича от базата абсолютно целият резултат от изпълнението на конструираната SQL заявка и се зарежда в оперативната памет. Това е най-бавната операция, която може да отнеме от няколко минути до няколко денонощия при по-големи бази, дори когато резултатът е малък по обем.

След като резултатът е зареден, потребителят може в интерактивен режим да извършва групиране на резултата по един или няколко атрибута. На фигура 2.9 е показано примерно групиране на пациентите по атрибутите *Град* и *Пол*.

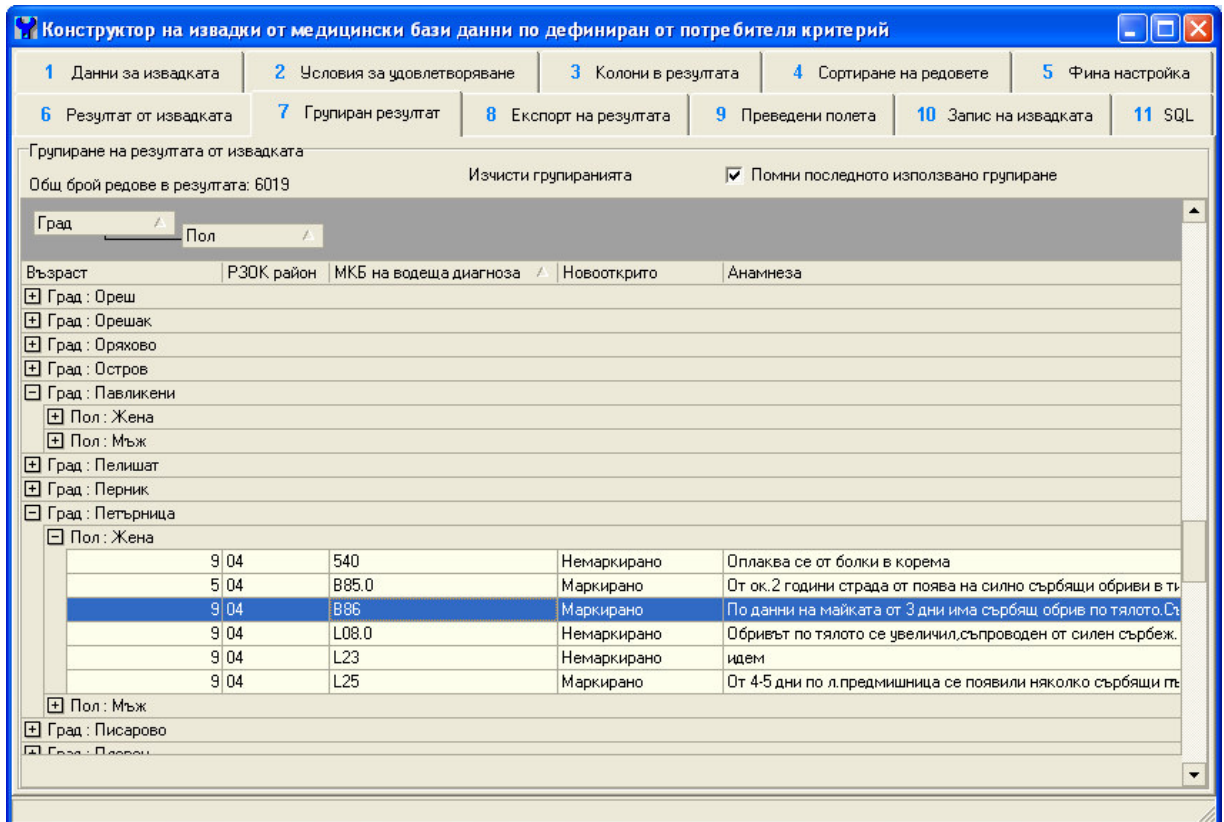

*Фигура 2.9. Стъпка 7 - „Групиран резултат"*

## 2.2.8. Стъпка 8 - „Експорт на резултата"

Тази стъпка позволява експорт на (евентуално групирания) резултат в няколко от най-популярните формати за обмен на данни. Те са:

- Във формат за зареждане от Визуализатора (данни и мета-данни в .data и .names формат)
- В клипборда в паметта, можете да го вмъкнете (с Paste) в произволна програма
- В чист текстов файл (.TXT), отваря се с произволен текстов редактор, например Notepad
- В хипертекстов файл (.HTML), отваря се с Интернет браузър, например Internet Explorer
- В електронна таблица (.XLS), отваря се с MS Excel

Освен файловият формат потребителят може да избере дали да се експортират всички данни или само тези редове, които той е избрал на стъпка 7 - „Групиран резултат". На фигура 2.10 е показана как изглежда тази стъпка в Конструктора.

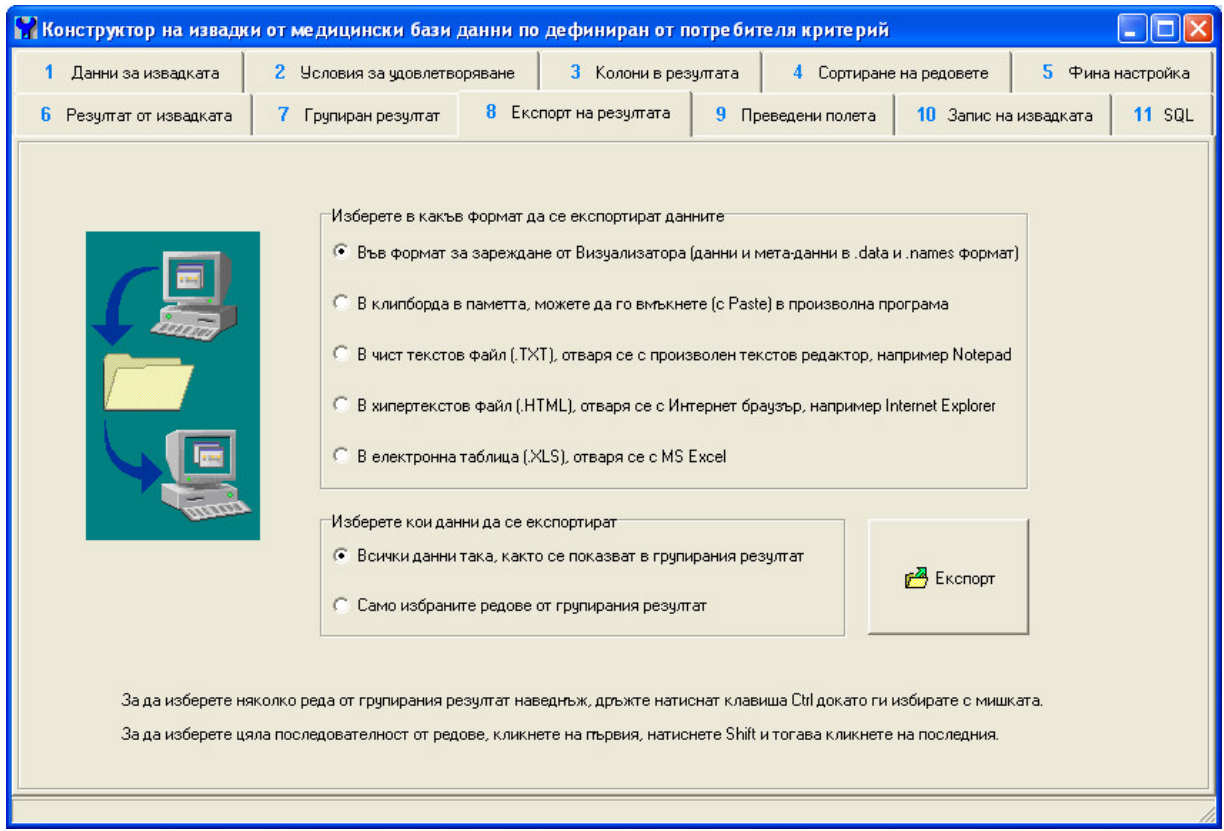

*Фигура 2.10. Стъпка 8 - „Експорт на резултата"*

## 2.7. Формат на данните и мета-данните

Форматът за зареждане от Визуализатора всъщност дефинира формат за два отделни файла.

- Единият файл съдържа мета-данни, т.е. описание на атрибутите от извадката и има разширение „.names", за да се запази подобието с разпространените бази от данни за DataMining.

- Другият файл съдържа самите данни от извадката и има разширение „.data". По същество той представлява текстов файл с разделител на данните (delimited text file), в който разделящ символ може да бъде табулация, запетая или друг такъв символ, който се очаква да не се среща в самите данни. Всеки ред от този текстов файл съдържа данните за един обект от извадката.

### 2.7.1. Формат на мета-данните

Споменатите по-горе .names файлове представляват валиден XML [XML, 2005] с много проста структура. Главният (root) елемент е с име „metadata" и xml атрибутите му съдържат незадължителни глобални параметри:

- description – описание на базата в свободен текст
- separator – разделителя между стойностите на атрибутите
- class – името на атрибута, който да се счита за клас/клъстер
- missingValue – съмволът, който означава, че липсва стойност за даден атрибут.

Единствените допустими под-елементи на елемента „metadata" са с име „attribute" и всеки от тях съдържа в xml атрибутите си настройки на съответния атрибут:

- name – задължително име на атрибута
- type – задължителен тип (непрекъснат или номинален)
- domain – допустими стойности за номинален атрибут. Този параметър няма смисъл за непрекъснати атрибути. Ако не е зададен domain, няма ограничение за допустимите стойности и те ще бъдат извлечени при прочитането на данните. Със задаване на domain, може да се постигне, както валидиране на данните, така и филтрирано прочитане, само на данни с определени сойности в съответния атрибут.
- skip – при стойност true, указва на програмата да не взима под внимание този атрибут.
- description – описание на атрибута в свободен текст

Формално, форматът на метаданните строго се подчинява на следната DTD схема (без да е необходимо да има експлицитно DOCTYPE обръщение към .dtd файл):

```
<!ELEMENT metadata (attribute+)>

<!ATTLIST metadata
  separator CDATA #REQUIRED
```

```
    missingValue CDATA #IMPLIED
    description CDATA #IMPLIED
    class CDATA #IMPLIED
>

<!ELEMENT attribute EMPTY>

<!ATTLIST attribute
    name ID #REQUIRED
    type (nominal | continuous) "nominal"
    domain CDATA #IMPLIED
    description CDATA #IMPLIED
    missingValue CDATA #IMPLIED
    skip (true | false) "false"
>
```

Ето един пример за мета-данни, демонстриращ използването на всички предвидени атрибути в XML формата:

```
<metadata
    separator=","
    missingValue="?"
    description="Syntax demo - animal attributes"
    class="Kind"
>
    <attribute
        name="ID"
        type="nominal"
        skip="true"
        description="Unique species name. Skiped during clasification."
    />
    <attribute
        name="Cover"
        type="nominal"
        domain="skin,fur,feathers,scales"
    />
    <attribute name="HasBlood" type="nominal" domain="yes,no" />
    <attribute
        name="Age"
        type="continuous"
        missingValue="-1"
        description="Age in years. Should be a positive integer."
    />
    <atttribute name="unknown" type="continuous" skip="true" />
    <attribute
        name="Weight"
        type="continuous"
        description="Floating point positive weight in kilos."
    />
    <attribute
        name="Kind"
        type="nominal"
        domain="mammal,bird,fish"
        description="The result of clasification/clusterization."
    />
</metadata>
```

## 2.7.2. Формат на данните

Споменатият по-горе „.data" файл по същество представлява текстов файл с разделител на данните (delimited text file). Разделящ символ може да бъде табулация, запетая или друг такъв символ, който се очаква да не се среща в самите данни и е указан от xml атрибута „separator" в мета-данните. Всеки ред от този текстов файл съдържа данните за един обект от извадката, като стойностите на атрибутите му са заградени с посочения разделящ символ. Броят и редът на атрибутите се определят от броя и реда на елементите „attribute" в мета-данните.

Пример на данни, съответстващи на дадения примерен .names файл с разделител запетая:

```
53, male, asympt, 140, 203, true, hyp, 155, true, 3.1, down, 0, rev, sick
60, male, notang, 140, 185, fal, hyp, 155, fal, 3, flat, 0, norm, sick
40, male, angina, 140, 199, fal, norm, 178, true, 1.4, up, 0, rev, buff
57, male, asympt, 165, 289, true, hyp, 124, fal, 1, flat, 3, rev, sick
60, male, asympt, 130, 253, fal, norm, 144, true, 1.4, up, 1, rev, sick
46, fem, asympt, 138, 243, fal, hyp, 152, true, 0, flat, 0, norm, buff
43, male, asympt, 110, 211, fal, norm, 161, fal, 0, up, 0, rev, buff
58, male, abnang, 120, 284, fal, hyp, 160, fal, 1.8, flat, 0, norm, sick
55, male, asympt, 160, 289, fal, hyp, 145, true, 0.8, flat, 1, rev, sick
41, male, abnang, 120, 157, fal, norm, 182, fal, 0, up, 0, norm, buff
52, male, notang, 172, 199, true, norm, 162, fal, 0.5, up, 0, rev, buff
62, fem, asympt, 138, 294, true, norm, 106, fal, 1.9, flat, 3, norm, sick
43, male, asympt, 120, 177, fal, hyp, 120, true, 2.5, flat, 0, rev, sick
47, male, asympt, 110, 275, fal, hyp, 118, true, 1, flat, 1, norm, sick
56, male, notang, 130, 256, true, hyp, 142, true, 0.6, flat, 1, fix, sick
74, fem, abnang, 120, 269, fal, hyp, 121, true, 0.2, up, 1, norm, buff
52, male, abnang, 120, 325, fal, norm, 172, fal, 0.2, up, 0, norm, buff
35, male, asympt, 126, 282, fal, hyp, 156, true, 0, up, 0, rev, sick
64, fem, asympt, 130, 303, fal, norm, 122, fal, 2, flat, 2, norm, buff
48, male, asympt, 122, 222, fal, hyp, 186, fal, 0, up, 0, norm, buff
58, male, asympt, 100, 234, fal, norm, 156, fal, 0.1, up, 1, rev, sick
51, fem, notang, 130, 256, fal, hyp, 149, fal, 0.5, up, 0, norm, buff
56, fem, asympt, 134, 409, fal, hyp, 150, true, 1.9, flat, 2, rev, sick
```

# 3. Проектиране на Визуализатора

Дотук проектирахме Конструктора на извадки с подходящ потребителски интерфейс, чрез който потребителят да работи лесно и удобно. В тази глава ще проектираме инструмента Визуализатор, който ще може да визуализира създадените извадки от Конструктора. Това ще позволи на потребителя да получи нагледна представа за структурата на данните. Допълнително ще проектираме и такива интерактивни функции, чрез които потребителят ще може да манипулира обектите от визуализацията. Само че, преди да можем да визуализираме данните от извадката, трябва внимателно да ги подготвим за обработка.

## *3.1. Подготовка на данните*

Съществуващите реални бази от данни съдържат зашумени, непълни и противоречиви данни, обикновено поради своя огромен размер, често достигащ до няколко гигабайта или дори повече. Непълнота в данните може да възникне по няколко причини. Някои от интересуващите ни атрибути могат да липсват, просто защото не са били въведени по време на създаване на съответния запис в базата от данни. Други стойности могат да липсват поради повреда в оборудването за тяхното събиране, например някой медицински апарат. Причините за наличието на зашумени данни или некоректни стойности на атрибутите могат да бъдат повреди в събиращото данни оборудване, човешки грешки при въвеждането на данни или грешки при тяхното получаване чрез комуникационни канали. Всички тези "замърсени" данни могат да объркат впоследствие ИЗД алгоритъма, водейки до ненадеждни и неправдоподобни резултати. По тази причина "изчистването" на данните е една важна стъпка от процеса на предварителната подготовка на данните преди да се прави опит за тяхната обработка. В този раздел ще представим базови методи за изчистване на данните и тяхното нормализиране.

### 3.1.1. Изчистване на данните от липсващи стойности

Съществуват множество различни подходи за справяне с липсващите стойности в данните. Следните подходи обикновено се приемат в ИЗД като базови:

    1. Игнориране на записа. Този метод обикновено се прилага, когато в записа липсва стойност на целевия атрибут (като се предполага, че задачата на ИЗД е класификацията) и не е много ефективен, освен в случаите, когато записът съдържа няколко атрибута с липсващи стойности. Методът води до лоши резултати в случаите, когато процентът на липсващите атрибутни стойности е значителен.

    2. Ръчно попълване на липсващите стойности. В общия случай този подход изисква много време и практически не е приложим за големи бази от данни с голям процент на липсващите атрибутни стойности.

    3. Използване на една глобална константа за заместване на липсващата стойност.

Всички липсващите стойности се заменят с една и съща константа, например, "?" (т.е. "неизвестно"). Някои версии на ИЗД алгоритмите (например, методи за класификация, използващи MVDM метрика за сходство) успешно построяват модели върху така "изчистени" бази от данни.

4. Запълване на липсващата стойност със стойността на средно аритметичното за даден атрибут. При този метод неизвестните стойности на един непрекъснат атрибут се заменят със средно аритметично на всички известни негови стойности, а неизвестните стойности на един номинален атрибут – с най-често срещаната известна негова стойност. Това е един доста разпространен метод за "изчистване" на данни, използван например при задачи на класификация, решавани чрез невронни мрежи.

5. Запълване на липсващата стойност със стойността на средно аритметичното за дадения атрибут с отчитане на класа. Този подход се използва при наличие на целевия атрибут. Неизвестната стойност на някой атрибут за всички записи от един и същ клас се попълва със средно аритметично на атрибута за същия клас, ако атрибутът е непрекъснат, или с най-често срещаната му стойност в класа, ако атрибутът е номинален.

6. Запълване на липсващата стойност с най-вероятната стойност на съответния атрибут. При този подход атрибутът с липсващите стойности се разглежда като целевия и задачата за попълване на тези стойности се третира като задача за класификация, ако атрибутът е номинален, или като регресия, ако той е непрекъснат. За тяхното решение могат да бъдат използвани такива методи като Бейсов класификатор, алгоритмите за най-близък съсед, класификационни или регресионни дървета и др.

Всички методи от 3 до 6 променят разпределението на данните и не отчитат възможните връзки между различните атрибути. От тази гледна точка последният метод използва най-много наличната информация за предсказване на липсващите стойности, като запазва връзките между атрибутите. Той обаче е доста по-сложен за изпълнение, тъй като изисква прилагане (и наличие) на различни класификационни или регресионни техники за моделиране.

### 3.1.2. Изчистване на данните от зашумени стойности

Шумът е случайна грешка или разсейване в стойността на измерваната величина. За непрекъснатите атрибути като измерител на разсейването (и по този начин показател за наличие на шум в данните) може да се използва коефициентът на вариация ($V_\%$), който изразява стандартното отклонение като процент от средното аритметично:

$$V_\%(X) = \frac{\sigma}{\tilde{x}} 100$$

където $\tilde{x}$ и $\sigma$ са съответно средното аритметично и стандартното отклонение на атрибута X.

Например, наличието на сравнително голямо разсейване (над 2%) може да бъде причинено от груби грешки при измерването или от наличието на сгрешени екстремални стойности. В първия случай "шумът" може да бъде намален чрез прилагане на методи за "изглаждане" (smoothing) на данни, докато във втория – чрез идентификация и премахване на съществуващите крайности (outliers).

Важно е да се отбележи, че голямо разсейване не винаги е признак на шум в данните – обикновено това се отнася само за еднородни данни, т.е. данни отнасящи към един и същ клас (например, когато става дума за стойността на температура на класа "Здрави хора"). Обратно, често това е признак на наличието в данни на отделни групи – клъстери или класове (например, голямото разсейване може да се получи при измерване на температура на здрави и болни хора).

Изчистване на зашумени данни може да стане чрез откриване и премахване на случайни грешки, които се проявяват като отклонение от общото поведение – екстремални стойности или крайности. Крайностите могат да бъдат определени чрез статистически анализ, например за такива могат да се считат стойностите, лежащи на разстояние 1.5 пъти по-голямо от между-квартилното разстояние от първия и третия квартил.

### 3.1.3. Изчистване на данните от противоречиви стойности

Противоречиви се наричат данни с грешна стойност на целевия атрибут (класа). За откриване на подобни противоречия се използват налични основни знания за проблемната област, например списък на допустими стойности на целевия атрибут - за откриване на грешки при въвеждане или неприемливи съкращения, известни функционални ограничения на стойности на някои атрибути от съответните класове (например атрибут „Брой левкоцити" не може да приема стойност „-20", нито стойността на атрибута „Тегло" за клас „Бебе" може да бъде по-голяма от 10 кг) и т.н.

Най-лесно се откриват синтактичните противоречия – това са случаи, когато в база от данни съществуват идентични записи, различаващи се само по стойността на целевия атрибут. Начинът за разрешаване на подобни противоречия е пълно премахване на всички противоречащи записи или само някои от тях. Това се прави според честотата на тяхното срещане и размера на самата база. Ако базата съдържа достатъчно записи от "спорните" класове, то подобни противоречиви записи могат да бъдат изтрити напълно. В противен случай окончателното решение може да се взима на база на определяне степента на сходство на спорните записи с всеки от противоречащите класове или чрез прилагане на някой класификационен алгоритъм, предсказващ стойността на класа за спорните записи.

### 3.1.4. Нормализация на данни

Една от често срещаните форми на трансформация на числови (непрекъснати) данни е тяхната нормализация – т.е. привеждането на стойностите на един атрибут в зададен диапазон. Нормализацията е особено полезна за създаване на класификационни модели с помощта на невронни мрежи или за методи, базирани на измерване на разстояния, какъвто е и избраният в тази дипломна работа метод за визуализация.

Съществуват различни методи за нормализация, като най-популярни от тях са: мини-максна нормализация, z-нормализация и нормализация чрез десетично мащабиране.

Мини-максната нормализация изпълнява линейна трансформация на оригиналните данни. Нека $min_A$ и $max_A$ са минималната и максималната стойности на атрибута A. Мини-максната нормализация изобразява всяка стойност v на A в стойността v' от диапазона [$new\_min_A$, $new\_max_A$] чрез трансформацията:

$$v' = \frac{v - min_A}{max_A - min_A}(new\_max_A - new\_min_A) + new\_min_A$$

Мини-максната нормализация запазва съотношенията между оригиналните стойности на данните. Тя може да се ползва за откриване на грешка в данни от типа "стойност извън диапазона", ако се окаже, че конкретната стойност на атрибута в някои от записи лежи извън оригиналния диапазон на атрибутните стойности .

Z-нормализацията (или нормализация с нулево средно аритметично) извършва нормализиране на данните на базата на средно аритметичното и стандартното отклонение на атрибута A. Стойността v на A се нормализира във v' чрез трансформацията:

$$v' = \frac{v - \tilde{A}}{\sigma_A}$$

където $\tilde{A}$ и $\sigma_A$ са, съответно, средно аритметично и стандартно отклонение на A. Този метод се използва, когато актуалните стойности на минимума и максимума на атрибута са неизвестни, или когато има екстремни стойности (ourliers), доминиращи в мини-максната нормализация.

Нормализация чрез десетично мащабиране изпълнява нормализацията чрез преместване на десетичната точка на стойностите на A. Броят на позициите при преместване на точката зависи от максимума на абсолютната стойност на A. Стойността v на A се нормализира във v' чрез трансформацията:

$$v' = \frac{v}{10^j}$$

където j е най-малкото цяло число, такова че max{|v'|} < 1.

Трябва да се има предвид, че след прилагането на нормализация данните се променят. За да бъдат нормализирани правилно и бъдещите данни, необходимо е да се пазят параметрите на използваната нормализация (например средното аритметично и стандартното отклонение при z-нормализацията).

## 3.2. Основа на визуализацията

Клъстеризацията е основен елемент в много подходи за извличане на закономерности от данни. За съжаление, резултатите от нея трудно могат да се възприемат от човек, когато данните са многомерни и разнородни. Това се случва когато обектите, които се клъстеризират, имат много характеристики, едни от които непрекъснати, а други – номинални и без наредба. Възниква необходимостта от визуално представяне на множеството от обекти и клъстерите, към които те принадлежат.

За основа на визуализацията в настоящата дипломна работа е избран алгоритъмът FastMap. Той е много удобен за визуализация и клъстерен анализ на многомерни данни. В този конкретен случай целта ни е оригиналните медицински данни да се проектират в двумерното пространство, като принадлежността към всеки клъстер да се маркира с фиксиран цвят по избрана цветова легенда. Други допълнителни ползи, които могат да се извлекат от една такава визуализация, са:

- Да се визуализира удобно резултатът от работата на някой алгоритъм за клъстеризация, за да се оцени от човек неговото качество.
- Да се прогнозира принадлежността на нов обект към вече съществуващите клъстери.
- Да се визуализират подходящо некласифицирани данни, за да може човек интуитивно да определи на колко клъстера ще е удачно да се клъстеризират данните от друг алгоритъм.

За основа на реализацията е използвано оригиналното описание на алгоритъма FastMap от авторите Zhexue Huang и Tao Lin. Описанието е взето от тяхната статия „A Visual Method of Cluster Validation with FastMap".

## 3.3. Алгоритъм FastMap

Алгоритъмът FastMap решава задачата за проектиране на *N* обекта, за които е известна *N x N* матрицата на взаимните разстояния, в *N* точки в k-мерно пространство по начин, запазващ (до голяма степен) съответствията в разстоянията между обектите. Решаването на задача за проектиране на *N* n-мерни обекта в *N* k-мерни обекта, където $k \leq n$, е частен случай на тази по-обща задача.

Вместо матрица на взаимните разстояния може да бъде използвана някаква мярка за разстояние между два обекта, дефинирана като функция D(A, B) := разстоянието между обектите A и B. В оригиналната версия на алгоритъма се използва Евклидово разстояние, но за нашите нужди ще дефинираме по-различни мерки за разстояние, които са съобразени с типовете на атрибутите от медицинските извадки.

При условие, че разполагаме с изчислени взаимни разстояния между обектите в оригиналното n-мерно пространство, намирането на проекциите на обектите в редуцирано k-мерно пространство се осъществява рекурсивно за k стъпки. На всяка стъпка се осъществяват следните действия:

1. Избират се два опорни (pivot) обекта $O_a$ и $O_b$. Линията, която минава през тях, се разглежда като първата ос на търсеното k-мерно пространство.

2. За всеки обект $O_i$ неговата координата $x_i$ по тази ос се изчислява по следната формула:

$$x_i = \frac{d_{a,i}^2 + d_{a,b}^2 - d_{b,i}^2}{2 d_{a,b}} \quad (1)$$

където $d_{a,b}$ е разстояние между $O_a$ и $O_b$, $d_{a,i}$ и $d_{b,i}$ са разстояния между $O_i$ и $O_a$ и $O_b$, съответно, които са вече изчислени (предварително или на предишната стъпка от рекурсията). Тази формула следва от косинусовата теорема и е илюстрирана на фигура 3.1.

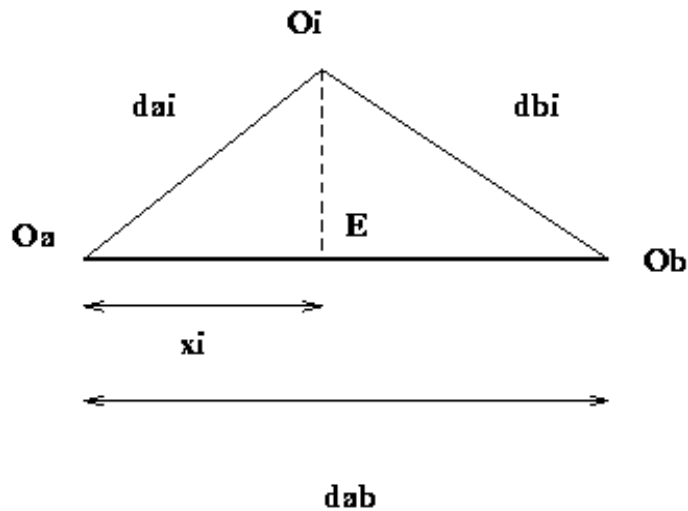

*Фигура 3.1. Прилагане на косинусовата теорема за проектиране върху правата $O_a O_b$*

3. За да бъдат изчислени координатите на обекти по следващата (в примера - втора) ос е необходимо предварително да бъдат преизчислени взаимните разстояния между всички обекти в намаленото (в случая n-1-мерно) пространство. Тези разстояния се изчисляват на базата на разстоянията в текущото пространство (в примера n-мерното) и координатите на обектите върху предишната (в случая – първата) ос по следната формула, която е следствие от Питагоровата теорема:

$$(d'_{i,j})^2 = (d_{i,j})^2 - (x_i - x_j)^2, \quad i,j = 1,\ldots,N \quad (2)$$

където $d'_{i,j}$ е разстояние между обектите $O_i$ и $O_j$ в редуцираното пространство, $d_{i,j}$ е разстоянието между обектите $O_i$ и $O_j$ в оригиналното (от предишната стъпка) пространство, а $x_i$ и $x_j$ са координати на съответните обекти върху първата (предишната) ос. По същество тази стъпка представлява проектиране на n-мерните обекти върху (n-1)-мерна хипер-равнина H, така както е илюстрирано на фигура 3.2.

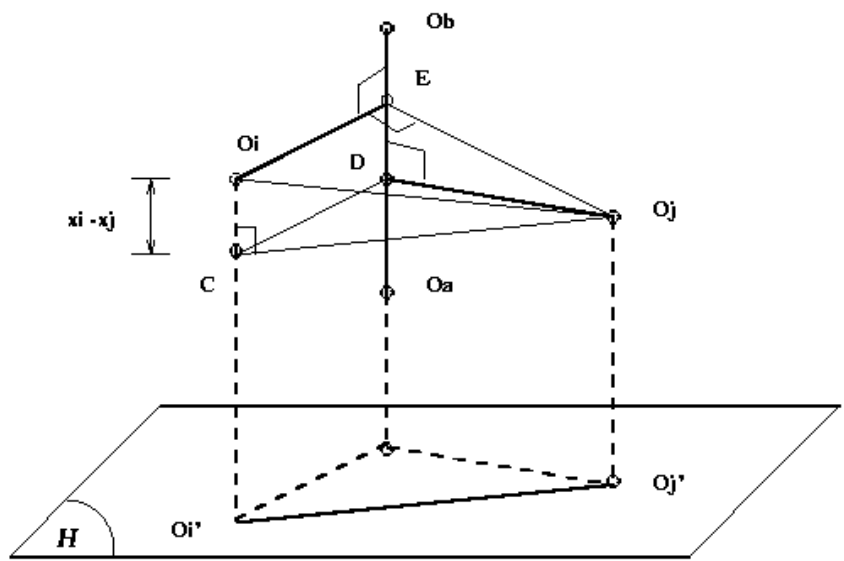

*Фигура 3.2. Проектиране върху хипер-равнина H, перпендикулярна на правата $O_a O_b$*

Описаните три точки се повтарят рекурсивно докато не бъдат изчислени координати на всички обекти върху k-тата координатна ос.

Опорните обекти $O_a$ и $O_b$ се избират по такъв начин, че да максимизират разстоянието $D(O_a, O_b)$. За да намали броя на необходимите за тази цел изчисления, които са $O(N^2)$, авторите използват следния евристичен алгоритъм със сложност $O(N)$ само за избор на опорни обекти:

---

**Алгоритъм избери_обекти(O, D())**

**Вход**
    Множество O от *N* обекта
    Функция за разстояние D()

**Начало**
    1. Избира се произволен обект, който се приема за втория опорен обект $O_b$.
    2. Като първи опорен обект $O_a$ се избира обектът, който се намира на най-голямо разстояние от $O_b$ съгласно избраната мярка за разстояние.
    3. В качеството на нов втори опорен обект $O_b$ се избира обект, който се намира на най-голямо разстояние от $O_a$.
    4. Стъпките 2 и 3 се повтарят зададено количество стъпки и получените накрая обекти се използват като опорни.

**Край**

---

В оригиналния алгоритъм броят на стъпките за избор на опорни обекти е равен на 5.

Целият алгоритъм FastMap може да се реализира чрез следния рекурсивен алгоритъм:

**Алгоритъм FastMap(k, D(), O)**

**Вход**
  Множество O от *N* обекта
  Функция за разстояние D()
  Желания брой на размерности k на целевото пространство

**Начало**
  Глобални променливи:
  **X**[] – *N* x k масив – в края на работата на алгоритъма i-ят ред на масива ще съдържа k-мерния образ на i-я обект
  **PA**[] - 2 x k масив – съхранява идентификатори на опорните обекти – по една двойка за всяко рекурсивно извикване.
  int j = 0 – указател на текущата колонка в X[]

  1) **АКО** k ≤ 0 **Край**
  **ИНАЧЕ**  j = j + 1

  2) /* Избор на опорни обекти */
  Нека $O_a$ и $O_b$ са резултат от работата на алгоритъма избери_обекти(O, D())
  по избор на опорни обекти, описан по-горе;

  3) /* Записване на идентификатори на опорните обекти */
  **PA**[1, j] = a; **PA**[2, j] = b;

  4) **АКО** $D(O_a, O_b)$ = 0
  X[i, j] = 0 за всички i и **Край** /* всички разстояния между обекти са 0 */

  5) /* проектиране на обекти върху линията $(O_a, O_b)$ */
  За всеки обект $O_i$
   Изчисли $x_i$ съгласно формула (1) и обнови глобалния масив:  **X**[i, j] = $x_i$

  6) /* проектиране на обекти върху хипер-равнина, перпендикулярна на линията $(O_a, O_b)$; функцията за разстояние D'() между двете проекции се задава с формула (2)
  **Рекурсивно извикване: FastMap(k – 1, D'(), O)**

**Край**

При всяко рекурсивно извикване алгоритъмът определя координатите на *N* обекта върху новата ос. Така i-тият обект се изобразява в точката Pi = (X[i, 1], X[i, 2], …, X[i, k]).

## 3.4. Дефиниране на разстоянието между обектите

Работата на алгоритъма FastMap разчита на непрекъснато преизчисляване на взаимните разстояния между обектите за всяка размерност на пространството. За тази цел трябва да сме в състояние да дефинираме подходяща функция D(A, B), която да дава числова мярка за разстоянието между всеки два обекта A и B. В тази секция ще разгледаме накратко как се изчислява разстоянието между обекти, описани с атрибути от различни типове данни. Ще разгледаме следните типове данни на атрибути – непрекъснати, двоични, номинални и атрибути с наредба.

### 3.4.1. Непрекъснати атрибути

Непрекъснатите или интервално-скалирани атрибути са непрекъснати числови стойности на измервания, направени съгласно една приблизително линейна скала. Типични примери са тегло, височина, температура и т.н. Трябва да се знае, че използваните мерни единици могат да окажат съществено влияние на резултатите от клъстерния анализ. Например, промяната от метри в инчове при измерване на височина може до доведе до получаване на съвсем различни клъстерни схеми. В общия случай, представянето на един атрибут в по-малки мерни единици води до по-голям диапазон на стойностите на този атрибут и, следователно, до по-голям ефект върху структурата на клъстерите. За да избегнем влияние от избора на мерните единици, данните трябва да бъдат стандартизирани (макар, че в някои приложения на някои атрибути могат да бъдат целенасочено присвоени по-големи тегла).

За да бъдат стандартизирани непрекъснатите атрибути, оригиналните измервания трябва да бъдат превърнати в безмерни единици. В лекцията, посветена на предварителната обработка на данните, вече разгледахме някои методи за стандартизация или нормализация на непрекъснатите данни. В клъстерния анализ често се прилага вече описаната z-нормализация, но с малки изменения:

$$z_{if} = \frac{x_{if} - \hat{x}_f}{s_f}$$

където $\hat{x}_f$ е средната стойност на атрибута f, а $s_f$ е средното абсолютно отклонение на този атрибут, изчислено по формулата:

$$s_f = \frac{1}{n} \sum_{i=1}^{n} | x_{if} - \hat{x}_f |$$

Средното абсолютно отклонение $s_f$ е по-устойчиво към екстремните стойности от "традиционното" стандартно отклонение $\sigma_f$, тъй като при неговото изчисляване отклоненията от средната стойност не се повдигат на квадрат и следователно, по този начин се намалява влиянието на екстремните стойности. От другата страна, използването на средното абсолютно отклонение не прави z-стойностите на екстремните стойности прекалено малки, т.е. тези крайности остават разпознаваеми, което е важно за анализа на крайностите.

Независимо от това, дали стойностите на непрекъснатите атрибута са стандартизирани или не, различието (или сходството) между два обекта, описвани чрез такива атрибути, се измерва с помощта на някоя мярка за разстояние. Един общ клас от такива мерки се задава от функцията за разстояние на Минковски:

$$d_L(i,j) = \left( \sum_{k=1}^{p} | x_{ik} - x_{jk} |^L \right)^{\frac{1}{L}}, L \geq 1$$

където i и j са два *p*-мерни обекта.

Най-често използваните разстояния в клъстерния анализ са Евклидово (при L=2) и абсолютното (при L=1). Претегленото разстояние на Минковски се получава когато всеки атрибут има различно тегло $w_f$:

$$d_{Lw}(i,j) = \left( \sum_{k=1}^{p} w_k | x_{ik} - x_{jk} |^L \right)^{\frac{1}{L}}, \sum_{k=1}^{p} w_k = 1; L \geq 1$$

### 3.4.2. Двоични атрибути

Един двоичен атрибут приема само две състояния: 0 или 1, където 0 означава, че атрибутът не присъства, а 1 – че присъства. Например, 1 за атрибут *Пушач* означава, че човекът пуши, а 0 - че не пуши.

При изчисляване на разстоянието между двоични атрибути се използват два подхода. Първият от тях се базира на изчисляване на така наречената таблица на случайностите (contingency table). Ако всички двоични атрибути се третират като имащи еднакво тегло, то за два обекта, описвани с *p* двоични атрибута, може да построи следната *2 x 2* таблица на случайностите:

**Обект *j***

|  | | 1 | 0 | **Сума** |
|---|---|---|---|---|
| **Обект *i*** | 1 | q | r | q + r |
| | 0 | s | t | s + t |
| | **Сума** | q + s | r + t | p |

където:
*q* е броят на общите атрибути (т.е. имащи стойност 1 и за двата обекта);
*r* е броят на атрибутите, присъстващи в обект i, но липсващи в обекта j;
*s* е броят на атрибутите, липсващи в обект i, но присъстващи в обекта j;
*t* е броят на атрибутите, липсващи и в двата обекта.

Общият брой на атрибутите е *p = q + r + s + t*.

Ще наричаме един двоичен атрибут симетричен, ако и двете му стойности имат еднакво тегло, т.е. няма никакво значение, кое от двете стойности на такъв атрибут да бъде кодирано с 1 и кое с 0. Например, атрибутът *Пол*, приемащ състояния мъж или жена, е симетричен. Сходството, оценявано на базата на симетрични атрибути, се нарича инвариантното сходство, тъй като резултатът не се променя при промяна на кодиране на някой от атрибутите. Оценката на разстоянието между два обекта, описвани със симетрични атрибути, най-често се прави чрез прост коефициент на съвпадение:

$$d(i,j) = \frac{r+s}{q+r+s+t} = \frac{r+s}{p}$$

Друга, често използвана мярка за разстояние между симетрични двоични атрибути е разстоянието по Хеминг (Hamming distance):

$$d_{Ham}(\mathbf{x},\mathbf{y}) = p - \sum_{i=1}^{p} x_i y_i - \sum_{i=1}^{p}(1-x_i)\cdot(1-y_i),$$

където:

$$\mathbf{x} = (x_1,...,x_p), \quad x_i \in \{0,1\}$$
$$\mathbf{y} = (y_1,...,y_p), \quad y_i \in \{0,1\}$$

Един двоичен атрибут се нарича асиметричен, ако неговите състояния имат различна важност (тегло). Например, положителен и отрицателен резултат на атрибута *"Тест за СПИН"* имат съвсем различно тегло! Обикновено за асиметричните атрибути чрез 1 се кодира състоянието, което се среща по-рядко (положителен – за *"Тест за СПИН"*). Следователно, за асиметрични атрибути положителното съвпадение (съвпадение на 1-ци) е по-важно от отрицателното. По тази причина, асиметричните атрибути се разглеждат не като двоични, а като "единични", т.е. имащи само едно състояние. Сходството, базирано върху асиметричните атрибути, се нарича не-инвариантно сходство. Най-известната мярка за изчисляване на разстояние между асиметрични атрибути е Жакардовия коефициент:

$$d(i,j) = \frac{r+s}{q+r+s}$$

Скаларното произведение е друг вариант на разстоянието по Хеминг, което често се прилага в случая на несиметрични атрибути:

$$d_{scalar}(\mathbf{x},\mathbf{y}) = p - \sum_{i=1}^{p} x_i y_i$$

Когато атрибутните стойности се кодират само с 0 и 1, при тази мярка само присъстващите признаци (т.е. 1-ци) допринасят за оценяването на сходството между обектите.

### 3.4.3. Номинални атрибути

Номиналният атрибут е едно обобщение на двоичния, при което атрибутът може да приема повече от едно състояния. Типичен пример на номинален атрибут е *Цвят*, който приема, например, състоянията *червен, жълт, зелен, розов* и *син*. Често за удобство М състояния на един номинален атрибут се кодират чрез цели числа – от 1 до

M, като не се предполага никаква наредба между тези числови стойности. Разликата между два обекта, описвани с номинални атрибути, може да се оцени чрез прост коефициент на съвпадение:

$$d(i, j) = \frac{p - m}{p}$$

където m е броят на съвпаденията – т.е. броят на атрибутите, които приемат една и съща стойност (състояние) и за двата обекта $i$ и $j$, а $p$ – общият брой на атрибутите. За да се увеличи ефектът от съвпадение е възможно на съвпадащите стойности да бъдат назначени различни тегла, или да бъдат назначени по-големи тегла на атрибутите, които могат да приемат по-голям брой възможни състояния.

### 3.4.4. Атрибути с наредба

Един дискретен атрибут с наредба (ordinal) е такъв номинален атрибут, на който неговите M състояния са подредени в определена, имаща смисъл наредба. Атрибутите с наредба се използват за описание на субективната оценка на такива величини, които не могат да бъдат измерени обективно. Например, професионалните звания често се представят като една наредба, от типа на *асистент, главен асистент, доцент, професор*. Непрекъснатият атрибут с наредба представлява един непрекъснат атрибут, измерен по неизвестна скала, т.е. при този атрибут важна е не конкретната му стойност, а относителният порядък на неговите стойности. Например, относителното подреждане в спорта (т.е. златен, сребърен, бронзов медал) често е по-важно от конкретния резултат. Атрибутите с наредба могат да бъдат получени чрез дискретизацията на непрекъснатия атрибут. Стойностите на атрибута с наредба могат да бъдат превърнати в рангове. Например, ако един атрибут с наредба $f$ има $M_f$ наредени състояния, то те могат да бъдат превърнати в ранговете $1, ..., M_f$.

При изчисляването на разстоянието между обектите, третирането на атрибутите с наредба е почти същото, като това на непрекъснатите атрибути. Ако $f$ е атрибут с наредба от $M_f$ наредени състояния, то разстоянието между два обекта по отношение на този атрибут се изчислява на следните 3 стъпки:

1. За всеки обект $i$ стойността $x_{if}$ на атрибута с наредба $f$ се заменя със съответния му ранг:

$$r_{if} \in \{1, ..., M_f\}$$

2. Тъй като различните атрибути с наредба могат да имат различен брой състояния, за да имат всички те едно и също тегло трябва техните стойности да бъдат стандартизирани, например в интервал [0.0, 1.0]. Това може да бъде постигнато чрез замяна на ранговете с техни нормализирани стойности:

$$z_{if} = \frac{r_{if} - 1}{M_f - 1}$$

3. Разстоянието между обектите се изчислява чрез използване на някоя от вече описаните мерки за разстояние, като атрибутът *f* вече се третира като непрекъснат със стойности $z_{if}$.

### *3.5. Проектиране на визуализацията*

След като вече разполагаме с методи за определяне на разстоянието между обекти, можем да приложим FastMap алгоритъма върху изходните данни от извадката и да получим представяне на обектите с намалена размерност. Остава да проектираме начин за визуализация на тези обекти.

Предлаганият начин за визуализация представлява двумерна (2D) визуализация на обектите, получени след изпълнение на FastMap алгоритъма. Всеки обект се визуализира чрез кръгче с определен диаметър. Тъй като данните са класифицирани, принадлежността им към съответен клас ще се използва за *автоматично оцветяване на обектите* от визуализацията в съответен на класа цвят по определена цветова схема. На фигура 3.3 е илюстрирана подобна визуализация.

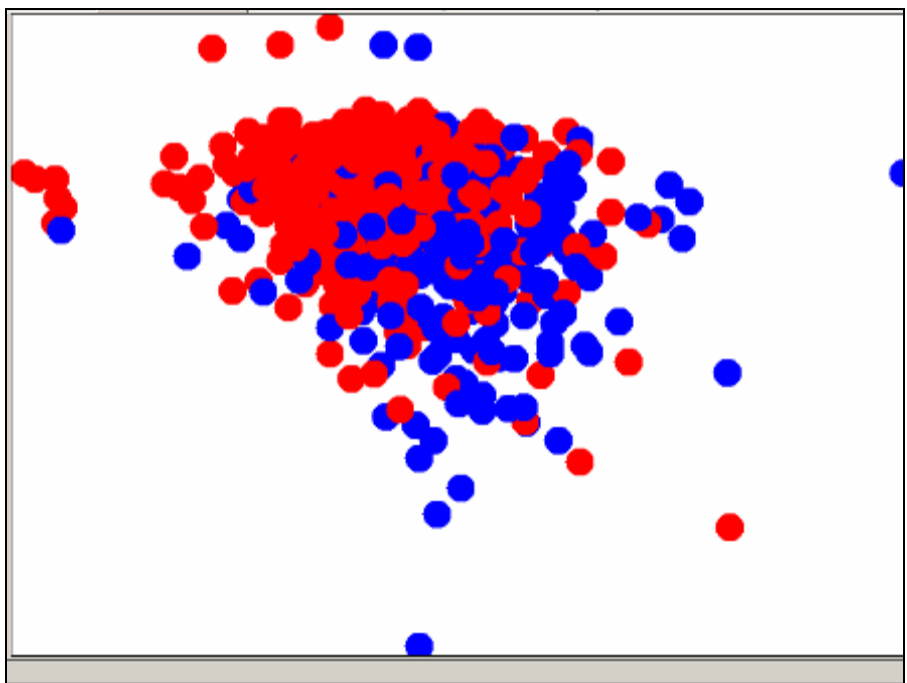

*Фигура 3.3. Визуализация на обекти с цветни кръгчета.*

Обичайните проблеми с 2D визуализациите по принцип са, че натрупването на много обекти на една и съща позиция не е достатъчно очевидно за потребителя. Това е така, тъй като най-често се визуализира само един от всичките обекти на дадена позиция. Ето защо за Визуализатора проектираме специален механизъм за по-добро визуализиране на застъпващи се обекти. Този механизъм е познат под името *Alpha-Blending* [Alpha-Blending, 2006] и се използва широко в тримерната компютърна графика. В резултат на неговото прилагане нагледно се вижда плътностното разпределение на обектите в двумерната проекция. На фигура 3.4 е илюстрирана същата визуализация като по-горе, но с използване на *Alpha-Blending* механизъм за изчертаване на обектите.

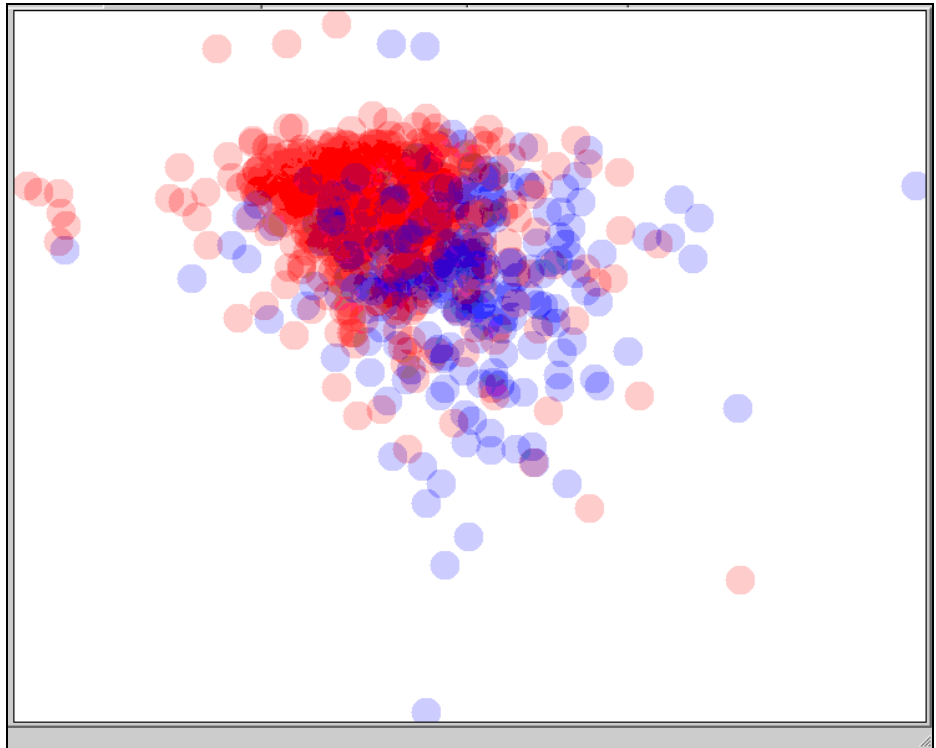

*Фигура 3.4. Визуализация на обекти с Alpha-Blending ефект.*

За да може потребителят да има *контрол над визуализацията* и работата на FastMap алгоритъма, проектираме възможност за настройването им интерактивно от Визуализатора. В таблица 3.1 са описани основните предвиждани настройки.

| Настройка | Описание |
|---|---|
| Размер на обектите | Определя размера на обектите от визуализацията. Тази настройка всъщност задава радиуса на кръгчетата (в брой пиксели), които представляват обектите |
| Alpha-Blending | Коефициент, определящ степента за прилагане на *Alpha-Blending* ефекта на смесване на цветовете на застъпващите се обекти. С тази настройка може да се контролира степента на смесване, което при различни видове извадки има различна оптимална стойност и зависи от степента на застъпване на обектите в двумерната проекция, както и от размера на обектите. |
| Брой стъпки за търсене на Pivot | Задава колко стъпки да се правят за откриване на двата най-отдалечени обекта на всяка стъпка от FastMap алгоритъма. Тъй като методът за намиране на тези 2 обекта е евристичен, то колкото повече стъпки се правят, толкова по-близо до истината ще бъде намерения резултат, но съответно толкова по-бавно ще работи алгоритъмът. |
| Z-Нормализация | Указва дали да се прави Z-нормализация на данните или не |

*Таблица 3.1. Настройки на визуализацията и FastMap алгоритъма.*

Както вече споменахме в секцията „Подготовка на данните", много често се налага потребителят да прави „изчистване" на данните от непълни, липсващи или противоречиви данни. Ето защо проектираме Визуализаторът да разполага с интерфейс, който да позволява на потребителя в табличен вид да *редактира данните от извадката*. Потребителят ще може да трие записи от извадката (например с липсващи стойности), да добавя нови записи ръчно, да редактира атрибутите на записите (например да дописва липсващи стойности или да прави корекции по тях). На фигура 3.5 е показан един табличен интерфейс за извършване на редактирането.

*Фигура 3.5. Табличен интерфейс за редакция на данните от извадката.*

Освен редактиране на самите данни, Визуализаторът ще осигурява удобен интерфейс за *редактиране на мета-данните*. Това включва редактиране на описанията на атрибутите, техните характеристики (тип, разделител, множество от допустими стойности и т.н.). Трябва да може чрез Визуализатора да се създават нови атрибути, да се премахват атрибути или да се маркират като несъществени, за да не влияят на алгоритъма за визуализация. На фигура 3.6 е показан примерен интерфейс за редактиране на мета-данните.

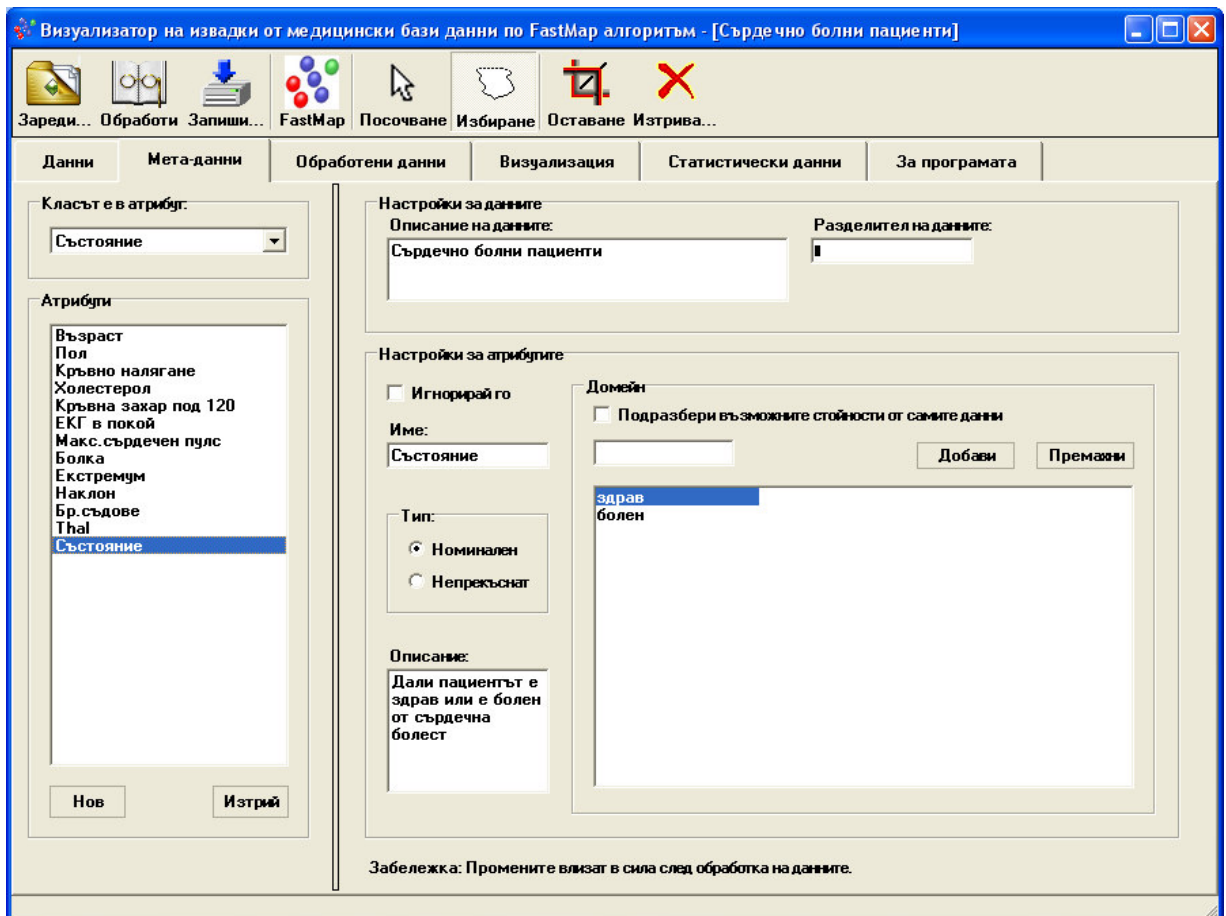

*Фигура 3.6. Примерен интерфейс за редактиране на мета-данните и избор на клас.*

Друга важна възможност на Визуализатора е *възможността за избор на клас*. Потребителят ще може да избира атрибут от наличните в мета-данните, който да изпълнява ролята на клас и по неговата стойност да се прави оцветяването на обектите в двумерната визуализация. Също така, потребителят ще може да укаже, че данните са некласифицирани и в този случай всички обекти от визуализацията ще бъдат с еднакъв цвят.

За да може потребителят лесно да проверява какво стои зад даден обект от визуализацията, проектираме *възможност за лесно разглеждане на обект* само чрез приближаване на курсора на мишката достатъчно близо до центъра му. Когато има такъв посочен с мишката обект, Визуализаторът автоматично ще показва данните на този обект в отделен панел, както е показано на фигура 3.7.

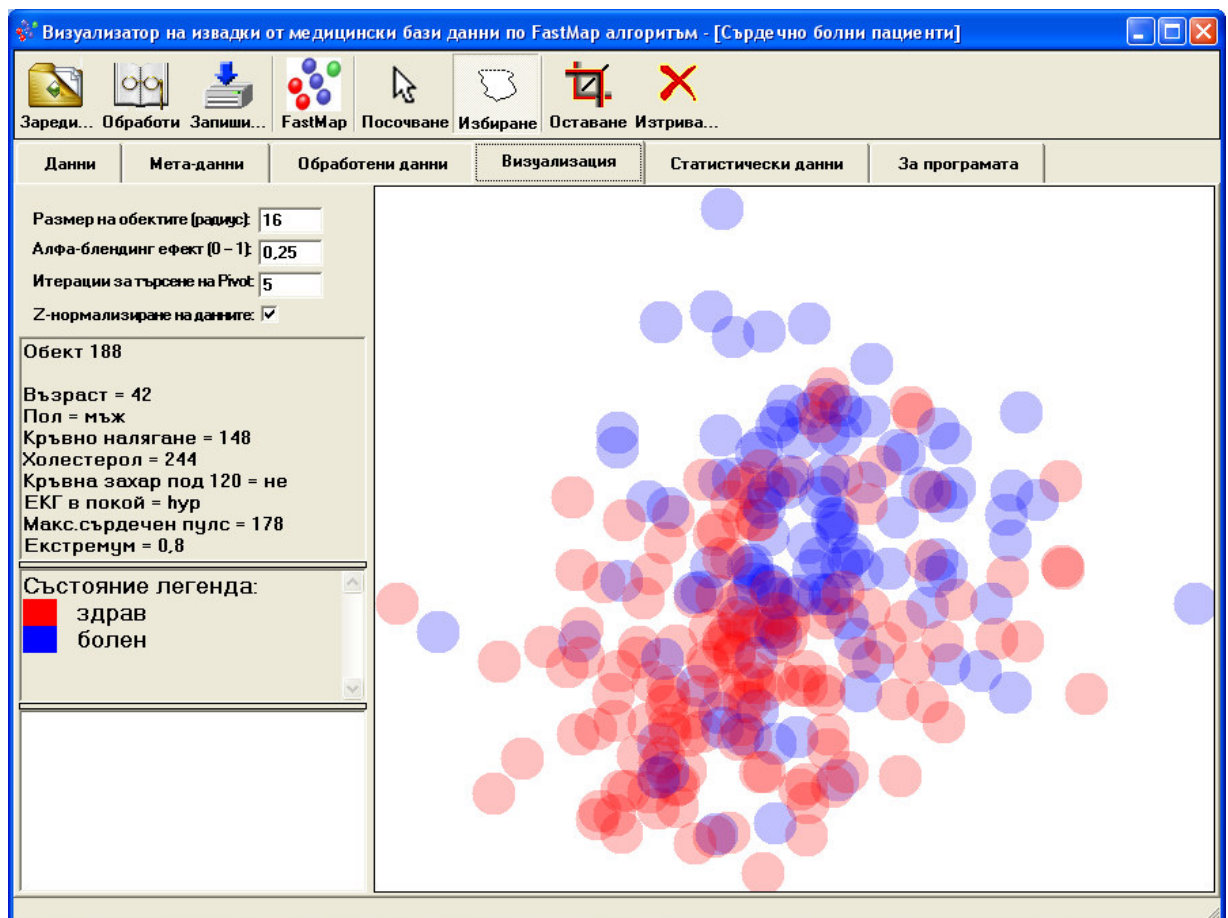

*Фигура 3.7. Примерен интерфейс за разглеждане атрибутите на посочения обекти от визуализацията.*

Възможността за *избор и манипулиране на обекти* от визуализацията има за цел да подобри интерактивността при работа с визуализацията, като същевременно позволи на потребителя да прави постъпкова визуализация на избрани от него подмножества от обекти. Избирането на един или повече обекти се извършва чрез задаване на свободна затворена крива в равнината с мишката. Визуализаторът автоматично прави избрани всички обекти, които попадат в запълнената вътрешност на зададената от потребителя крива. Към множеството от текущо избраните обекти могат със същата операция да се добавят още обекти по избор на потребителя. На фигура 3.8 е показан примерен интерфейс за манипулиране на обектите от визуализацията.

Върху избраните обекти се предоставят две възможни операции: *Оставане (Crop)* и *Изтриване (Delete)*. Оставането премахва всички неизбрани обекти, като следващото стартиране на FastMap алгоритъма работи само с останалите (избраните) обекти. Тази операция се използва, когато потребителят иска да види някаква интересуваща го част от извадката в по-разгърнат вид. *Изтриването* прави обратното на *Оставането* и най-често се използва за отстраняване на крайностите в извадката (outliers) и повторна визуализация. Това често пъти е резултатно, тъй като именно крайностите е много вероятно да са били избрани за опорни (pivot) елементи на стъпката от алгоритъма, на която се прави търсене на двата най-далечни обекта.

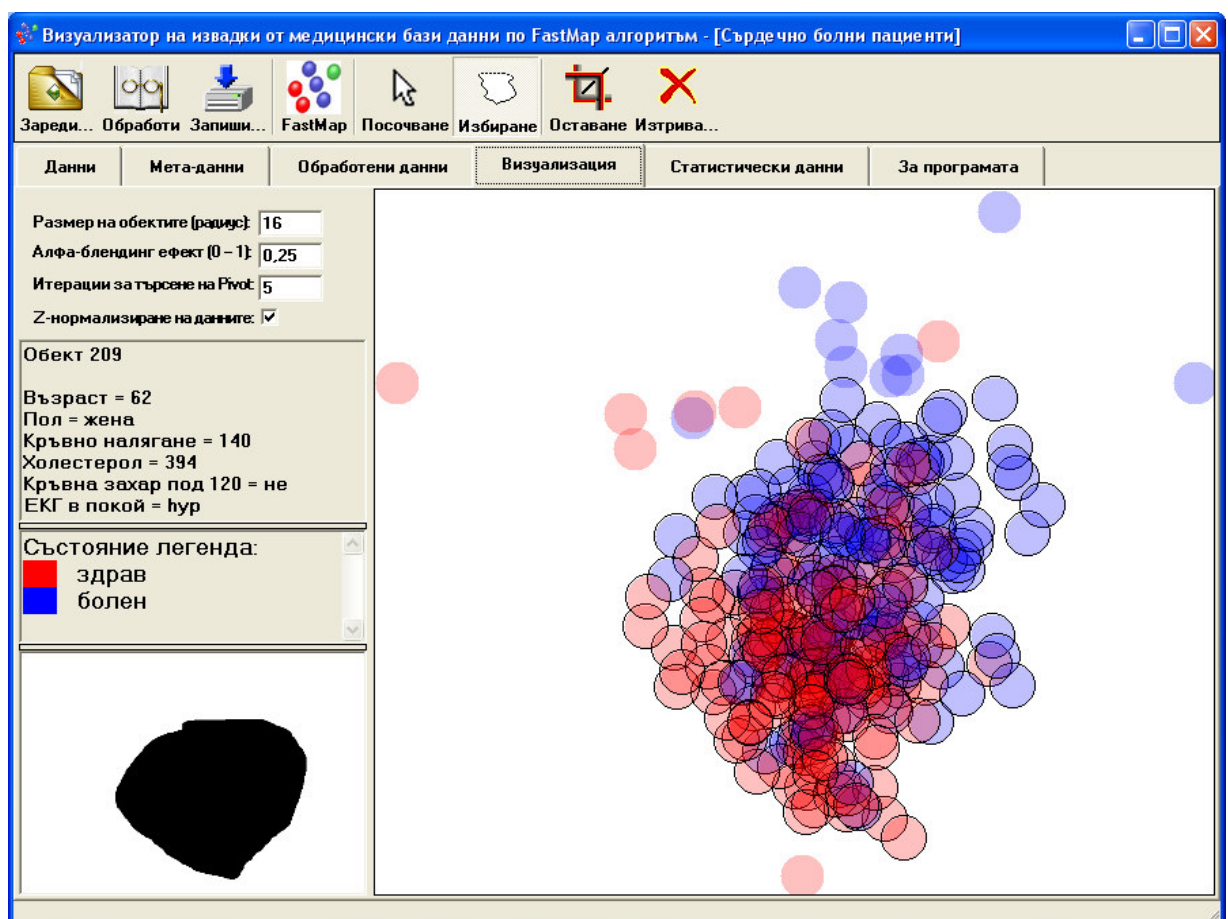

*Фигура 3.8. Примерен интерфейс за манипулиране на обектите от визуализацията.*

На последно място, Визуализаторът предоставя *възможност за записване/експорт/* на обработените данни във външни файлове, за да могат те да се подадат на друга външна програма за повторна клъстеризация или друга обработка. Така цикълът на ИЗД се затваря и потребителят има възможност да прави много итерации през инструментите с все по-отбрани данни.

### 3.6. Оценка на качеството на визуализация

Вече споменахме, че Визуализаторът може да се използва за нагледна проверка на резултата от работата на някой ИЗД алгоритъм. В частност, Визуализаторът може да послужи за оценка на качеството на алгоритъм за клъстеризация.

Качеството на клъстеризация може да бъде оценено не само от човек, но и чрез други, формални мерки. Две от тях са квадратът на претегления среден радиус на клъстеризацията ($\bar{R}$) и квадратът на претегления среден диаметър на клъстеризацията (D), които се определят по следните формули:

$$\overline{R} = \frac{\sum_{i=1}^{k} n_i R_i^2}{\sum_{i=1}^{k} n_i} \quad u \quad D = \frac{\sum_{i=1}^{k} n_i(n_i-1)D_i^2}{\sum_{i=1}^{k} n_i(n_i-1)}$$

където:

$k$ – е броят на клъстерите

$n_i$ – е броят на обектите в клъстер $i$

$R_i$ и $D_i$ са, съответно, радиус и диаметър на $i$-я клъстер, изчислени по формулата:

$$R_i = \sqrt{\frac{\sum_{j=1}^{n_i} d^2(\mathbf{x}_j, \mathbf{m}_i)}{n_i}}, \quad \mathbf{x}_j \in C_i$$

$$D_i = \sqrt{\frac{\sum_{k=1}^{n_i}\sum_{j=1}^{n_i} d^2(\mathbf{x}_k, \mathbf{x}_j)}{n_i(n_i-1)}}, \quad \mathbf{x}_j \in C_i, \mathbf{x}_k \in C_i$$

където:

$\mathbf{m}_i$ е центроид на съответния клъстер $C_i$.

Визуализаторът пресмята диаметрите на всеки един клъстер преди и след FastMap трансформацията на обектите, като показва резултатите в две кръгови диаграми (pie charts). Така лесно може да се види доколко изображението на N-мерните обекти в 2-мерното пространство, извършено от FastMap, е успяло да запази взаимното пространствено разпределение на обектите. Това до някъде говори за *качеството на самата визуализация*, т.е. колко е степента на изкривяване на визуализацията поради редуцирането на размерността.

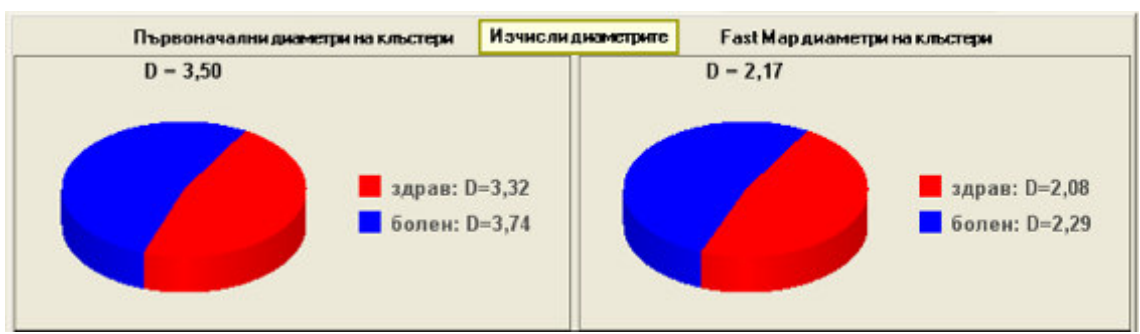

Освен пресмятането на споменатите диаметри, Визуализаторът изчислява и други статистически характеристика за всеки атрибут според неговия тип. За *номиналните атрибути* се изчислява броят срещания на всяка от допустимите стойности в домейна на атрибута и резултатите се показват в кръгова диаграма с легенда.

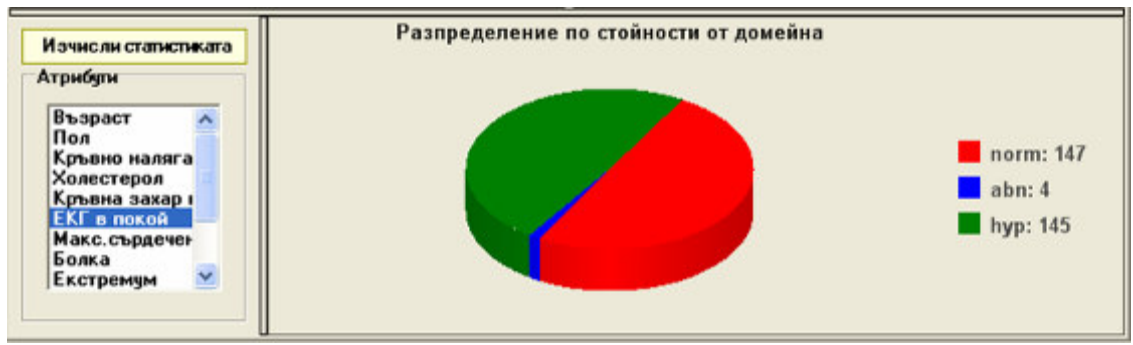

За *непрекъснатите атрибути* се изчисляват други две основни статистически характеристики. Това са средната стойност и стандартното отклонение на стойностите на атрибута, които се показват в стълбова диаграма (bar chart).

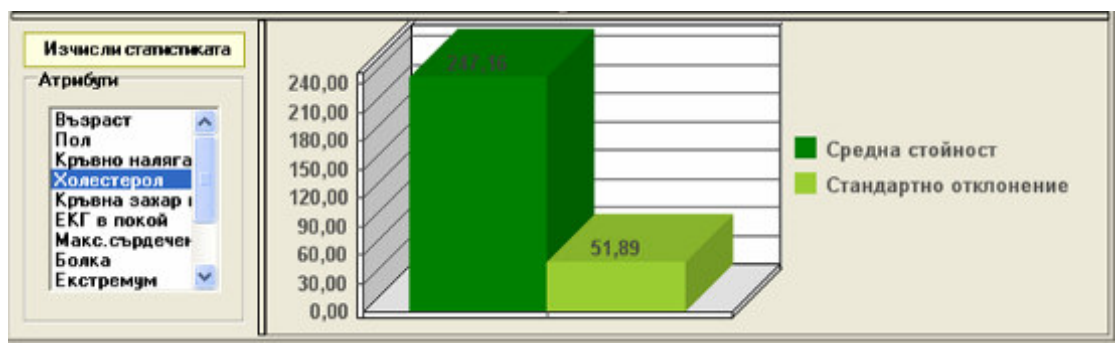

# 4. Реализация

Този раздел описва конкретната реализация на двата инструмента, използваните езици за програмиране и софтуерни технологии за конструиране и визуализация на извадки.

Едни от най-тежките за удовлетворяване изисквания към настоящата дипломна работа са свързани главно с интерфейса и това доколко той е интуитивен и лесен за използване от широк кръг потребители. Ето защо изборът на средства за реализация е изключително важен за постигане на желаните резултати.

## *4.1. Избор на средства за реализацията*

Може да се каже, че дипломната работа е реализирана като хетерогенна софтуерна система с развързана (loosely coupled) архитектура, в която двата основни модула "Конструктор" и „Визуализатор" обменят информация чрез специално разработени формати на данните и мета-данните. Тази архитектура е възприета, тъй като дава възможност за самостоятелно независимо използване на двата инструмента, както и „отваря" системата за съвместно използване с други подобни инструменти. Например, напълно възможно е друга ИЗД система да подаде данни направо към Визуализатора за разглеждане от потребителя. Възможен е и другият вариант: Извадка, създадена с Конструктора, да се подаде на някой друг ИЗД инструмент за обработка, например за клъстеризация. По време на тестването и настройването на дипломната работа и двата варианта бяха изпробвани и дадоха задоволителни резултати.

Като платформа за реализацията на Конструктора е избрана средата Borland Delphi 6.0 [Delphi, 2006], а за достъп до базата от данни се използват DBExpress компоненти [DBExpress, 2006]. Като RDBMS сървър за бази от данни се използва Borland InterBase 6.0 [InterBase, 2006]. Изборът на този език за програмиране и сървър за бази от данни е направен заради по-лесната интеграция с програмния продукт „Хипократ" [Хипократ, 2006]. Инструментът „Конструктор" от настоящата дипломна работа успешно беше интегриран в програма „Хипократ" и вече повече от 1 година потребителите на програмата го използват за създаване на гъвкави извадки от своите бази с медицински данни.

Като платформа за реализацията на Визуализатора е избрана средата Microsoft Visual Studio .NET 2003 [Visual Studio, 2006], а за език за програмиране – езикът C# от платформата .NET Framework. Изборът на този език за програмиране е направен заради наличието на мощни библиотеки за двумерна визуализация, които са продукт на Microsoft и имат много добра интеграция с операционната система Microsoft Windows, затова постигат голяма графична производителност. За визуализация е използвана библиотеката "Drawing 2D" [Drawing2D, 2006], която направи възможно използването на проектирания Alpha-Blending ефект за по-добра визуализация на плътностното разпределение на обектите.

## *4.2. Реализация на Конструктора*

Разглеждаме реализацията на Конструктора от две страни: външна реализация на потребителския интерфейс и вътрешна реализация на фунционалността.

### 4.2.1. Реализация на потребителския интерфейс

Външният вид на Конструктора е изцяло съобразен с проектираните стъпки на процеса на конструиране. Всяка стъпка от процеса е отделена в самостоятелна страница от PageControl компонента и е номерирана, за удобство на потребителите. Предпочетен е този начин за изграждане на интерфейса пред последователно изреждане на стъпките (Wizard style), тъй като той позволява във всеки един момент пряк достъп до произволна стъпка и по този начин много по-бърза и удобна корекция на настойките за напредналите потребители.

Помислено е и за начинаещите потребители. За тях е предвиден по-опростен режим, в който функционалността е сведена до един достатъчен минимум, позволяващ конструиране на елементарни извадки от базата с много опростен интерфейс.

Външният вид на всяка стъпка на конструиране е показан с типичен общ изглед в Таблица 4.1.

*Таблица 4.1. Реализация на потребителския интерфейс на всяка стъпка.*

| Стъпка | Общ изглед |
|---|---|
| 1. Данни за извадката | 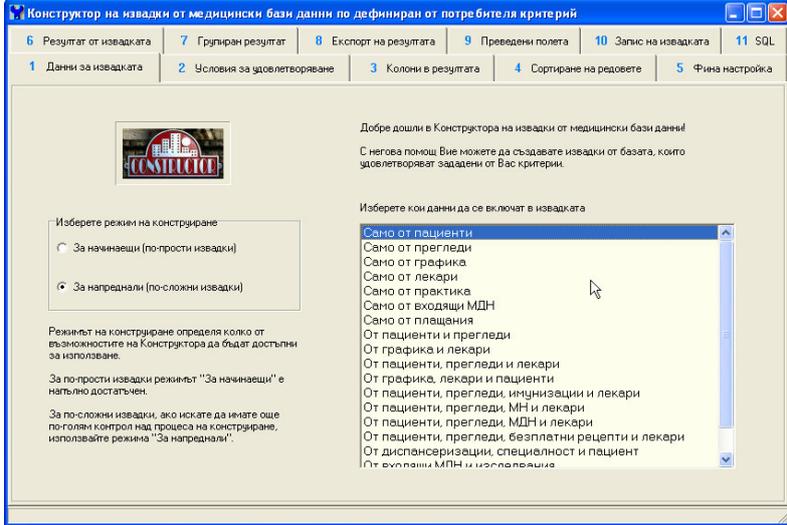 |

| Стъпка | Общ изглед |
|---|---|
| 2. Условия за удовлетворяване | |
| 3. Колони в резултата | |
| 4. Сортиране на редовете | |

| Стъпка | Общ изглед |
|---|---|
| 5. Фина настройка | |
| 6. Резултат от извадката | |
| 7. Групиран резултат | |

| Стъпка | Общ изглед |
|---|---|
| 8. Експорт на резултата | 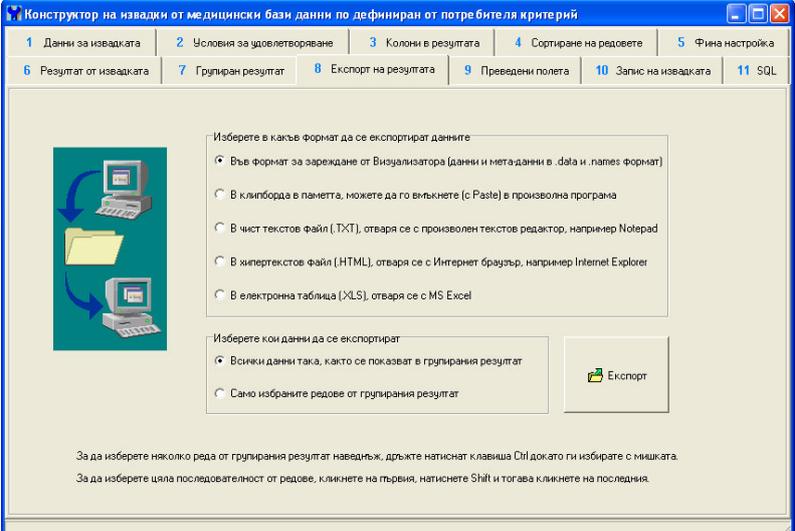 |

*Таблица 4.1. Реализация на потребителския интерфейс на всяка стъпка.*

### 4.2.2. Реализация на функционалността

Конструкторът е реализиран изцяло върху обектно-ориентираната платформа Borland Delphi 6.0 с използване на компоненти DBExpress за достъп до базата от данни. Ще направим кратко описание на реализацията на функционалността на програмата, разделено по нейните основни класове:

*Konstruktor.dpr*
От този модул започва изпълнението на програмата. В него се създава главната форма (frmConstructor), която съдържа целия потребителски интерфейс.

*Constructor_du*
Този клас представлява т.нар. „Модул за данни" (TdataModule), който съдържа Delphi класове, отговарящи за връзката с базата от данни и осъществяващи изпълнение на конструираните SQL заявки и съответно обратно изтегляне на върнатия резултат.
В този клас се намира преобразуването на имената на елементите от базата (таблици, полета, стойности) в удобни за потребителя имена на български език. Тази информация се пази в специални външни .map файлове, в които се дефинира съответствието между елемент от схемата на базата и наименование в интерфейса на Конструктора.

*ConstructedQuery*
Този клас капсулира една конструирана SQL заявка със всички нейни клаузи: select, from, join, where, unique, order by и group by. Този клас е отговорен за реализирането на „умното свързване" на условията и за прилагане на фините настройки на изпълнението (например Unique клаузата).

*Constructor_un*
Този модул съдържа формите за потребителския интерфейс, отделните стъпки в конструирането на една заявка и контролите, с които потребителят работи.

*SQLFiltr*
Този клас реализира функционалността на стъпка 2 – „Условия за удовлетворяване" от Конструктора. Той отговаря за създаване на динамичен интерфейс за задаване на условия, съобразен с информацията за типовете на атрибутите от схемата на базата.

*UtilListBox*
Този помощен модул се използва за реализиране на потребителския интерфейс на стъпките „Сортиране" и „Избор на полета", като функционалността му се споделя от двете стъпки в еднаквата си част.

*progress_un*
Този клас създава и показва при необходимост анимираното съобщение „Моля, изчакайте". Колкото и да е елементарна неговата функционалност, тя е съвсем необходима, тъй като потребителите трябва да имат обратна връзка от програмата по време на изпълнение на (понякога доста бавните) заявки към базата.

*Utilities*
Този клас съдържа спомагателни процедури за използване от цялата програма.

### *4.3. Реализация на Визуализатора*

Първоначално Визуализаторът беше реализиран точно така, както беше замислен при проектирането му. Оказа се обаче, че непосредственото боравене със стойностите на номиналните атрибути забавя значително работата на FastMap алгоритъма, тъй като се налага преобразуване на данните за всяко пресмятане на разстояние. Практически се оказа, че така реализираният инструмент не беше способен да работи с повече от няколко хиляди обекта, защото ставаше много бавен.

За да се реши проблемът с бързодействието, беше добавена една предварителна обработка на данните, целяща еднократно в началото да се извършат всички възможни преобразования, така че после да не се губи време за това по време на работата на FastMap алгоритъма. По време на тази предварителна обработка се прави *индексиране на номиналните атрибути*, тоест прави се преобразуване на номиналните атрибут в индексни стойности, които се използват вместо оригиналните стойности за по-бързо пресмятане на разстоянията. В допълнение, ако потребителят е задал да се извършва z-нормализация на данните, това се прави именно в този момент при предварителната обработка на данните.

Ще разгледаме реализацията на Визуализатора отново от две страни: външна реализация на потребителския интерфейс и вътрешна реализация на фунционалността.

### 4.3.1. Реализация на потребителския интерфейс

Външният вид на Визуализатора е изцяло съобразен с проектираните функции на процеса на визуализация. Отделните етапи от процеса са отделени в самостоятелни страници от PageControl компонента, а в горната част на основния прозорец е изведен Toolbar с бутони за извършване на основните операции. Външният вид на Визуализатора след стартиране е показан на фигура 4.1.

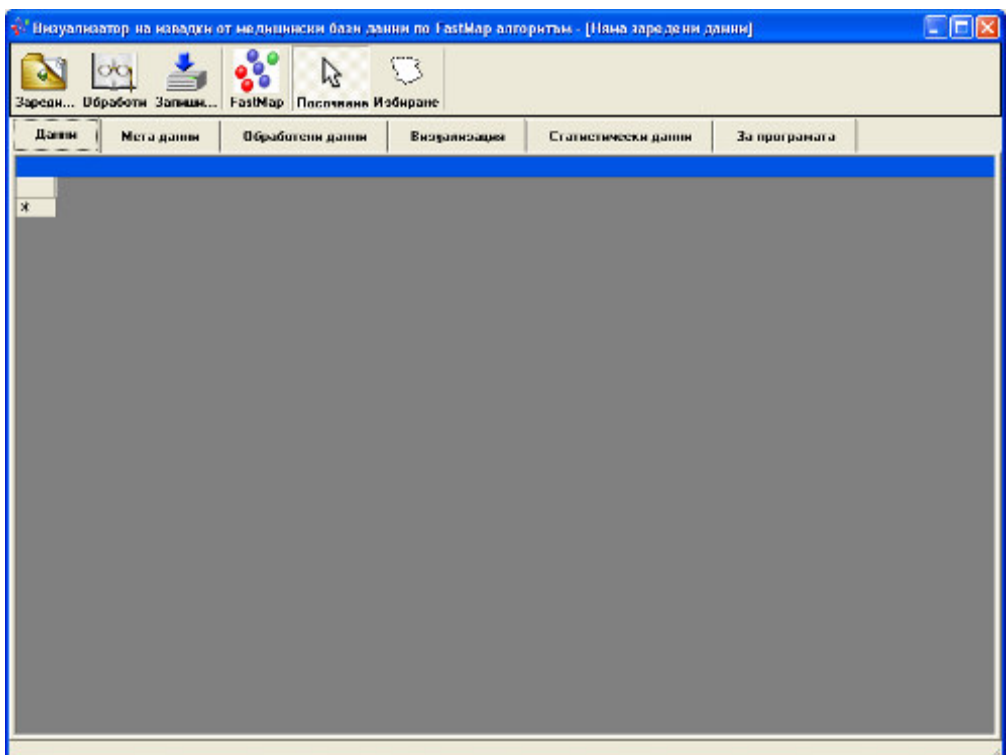

*Фигура 4.1. Външен вид на Визуализатора*

Работата с програмата протича обикновено в следната последователност: зареждане на данните и мета-данните, редактирането им, обработка на данните, визуализация, операции над визуализациятата, събиране на статистически данни. Накратко ще опишем един примерен сценарий за работа с програмата. Примерни екрани, демонстриращи работата със всяка страница от потребителския интерфейс, са дадени в Таблица 4.2.

След стартиране на програмата чрез бутона "Зареди..." се избира извадка за отваряне. Визуализаторът зарежда данните от .data файла и мета-данните от .names файла. Данните и мета-данните се показват в едноименни страници на екрана.

В страницата „Данни" потребителят има пълна свобода да редактира всички стойности. Това включва промяна, добавяне и изтриване на обекти (редове). Триене на цял ред от данните става като той първо се маркира целия с мишката и се натисне клавиш Del.

В страницата „Мета-данни" потребителят може да разглежда и да редактира описанието на атрибутите. Това включва избор на тип за всеки атрибут (номинален или непрекъснат), задаване на валидни стойности за домейна на номиналните атрибути, изключване на атрибути, които потребителят не иска да влияят на визуализацията и т.н. Визуализаторът може сам да намери валидните стойности за домейна на базата на срещаните в данните стойности. Важен момент в тази страница е изборът на атрибут, който играе ролята на клас. Този избор определя оцветяването на обектите от визуализацията. В частност, може да се избере, че липсва класов атрибут, като в този случай всички обекти от визуализацията са с един цвят.

Станицата "Обработени данни" показва преобразуваните данни в индексен вид, който се използва вътрешно от програмата за оптимизация на производителността. Тук се забелязва, че всички атрибути, маркирани от потребителя за пропускане в страницата „Мета-данни", са придобили стойност (null). В тази страница всички данни са преобразувани в числови и са z-нормализирани (ако потребителят е поискал това). Ако потребителят и променил необработените данни на ръка, трябва да се натисне бутонът "Обработи", за да се отразят промените в обработените данни.

Страницата „Визуализация" съдържа следните елементи:

- Двумерната визуализация, резултат от работата на FastMap алгоритъма
- Легенда за назначените цветове на всяка стойност на атрибута за клас
- Опциите за настройка на визуализацията и работата на алгоритъма
- Информация за текущо посочения с мишката обект, която съдържа двойки „Атрибут – стойност" за съответния посочен обект и уникален номер на обекта, което позволява той да бъде еднозначно идентифициран при различни визуализации.
- Графичната маска за избор на обекти, получена от натрупване на изчертаните с мишката затворени криви. Това е черно-бяло изображение, като черните участъци показват областите, в които всички намиращи се обекти ще бъдат избрани.

В страницата „Визуализация" има 2 режима на работа: *Посочване* и *Избиране*. В режим *Посочване* курсорът на мишката има формата на кръстче и при доближаване на малко разстояние до някой обект от визуализацията в левия информационен панел се изписва информацията за него. В режим *Избиране* са налични двете операции *Оставане* и *Изтриване*, които работят по начина, проектиран в предната глава. Курсорът в този режим е под формата на стрелка и позволява изчертаване на затворени криви, които се отразяват веднага на графичната маска в долния ляв ъгъл и се извършва избиране на попадащите в маската обекти. Избраните обекти се очертават с черни окръжности във визуализацията.

Петата страница („Статистически данни") показва изчисленията на характеристики като диаметри на клъстерите, разпределение на стойностите и други показатели, които дават информация за качеството на визуализация и друг синтезиран поглед върху разпределението на данните. За различните типове атрибути се показват различни характеристики по начина, описан в главата Проектиране.

Шестата страница („За програмата") съдържа информация за дипломната работа.

*Таблица 4.2. Примерни екрани от Визуализатора, демонстриращи работата със всяка страница от потребителския интерфейс.*

| Страница | Общ изглед |
|---|---|
| Данни | |
| Мета-данни | |

| Страница | Общ изглед |
|---|---|
| Обработени данни | 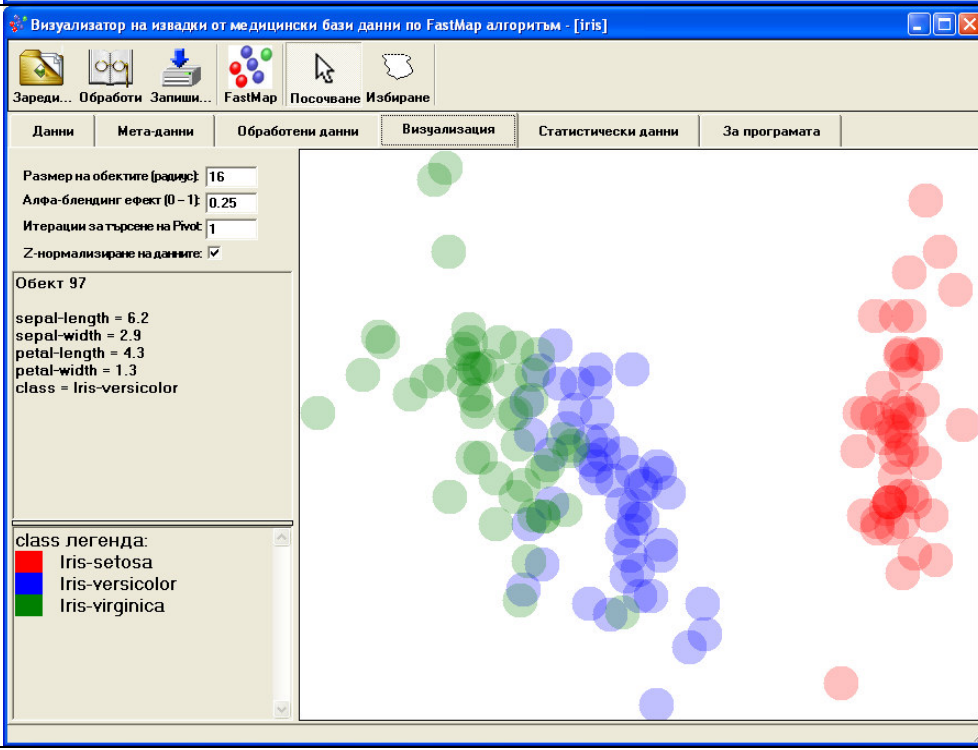 |
| Визуализация в режим *Посочване* | |

| Страница | Общ изглед |
|---|---|
| Визуализация в режим *Избиране* | |
| Статистически данни за номинален атрибут | |

| Страница | Общ изглед |
|---|---|
| Статисти-чески данни за непрекъснат атрибут | |

### 4.3.2. Реализация на функционалността

Визуализаторът е реализиран изцяло върху обектно-ориентираната платформа Microsoft .NET с използване на езика C# и библиотеката Drawing2D. Ще направим кратко описание на реализацията на функционалността на програмата, разделено по нейните основни класове:

*Global*
От този клас започва изпълнението на програмата. В него се създава главната форма (прозорец), която съдържа целия потребителски интерфейс.

*MainForm / MetadataForm / StatisticsForm*
Потребителският интерфейс на проекта е реализиран в тези три модула. Приложението използва интерфейс тип SDI (Single Document Interface), като модулите са прикачени в отделни страници (TabSheets). От главната форма се създават обектите, които представят базата от данни (mDatabase), алгоритъма (mFastMap), визуализацията (mVisualization) и настройките (mOptions). Повечето методи в тези класове касаят визуализацията на данните.

*FastMap*
Този клас реализира основния алгоритъм за получаване на 2D координати на обектите по начина, описан при проектирането на Визуализатора. Алгоритъмът се базира на непрекъснато преизчисляване на разстоянията между обектите. Функцията за разстояние е рекурсивна, използва базовите разстояния и изчислените до момента изменения в тях, породени от проектирането. Разстоянията се пресмятат в момента в който потрябват, с което се избягва квадратичната сложност по поддържането на матрица на разстоянията. Базовото разстояние се смята като Евклидово разстояние за непрекъснатите атрибути и като разстояние по Хеминг за номинални атрибути (0 за съвпадащи и 1 за различаващи се стойности).

*Options*
Всички параметри, които подлежат на настройка са изведени в отделен модул *Options*. Когато в опциите е избрана нормализация, за изчисляване на разстоянията се използват Z-нормализираните стойности на непрекъснатите атрибути.

*Visualization*
Реализира вътрешното представяне на елементите, които се визуализират (като VirtualSpace – списък от VirtualObjects) Поддържа текущото подмножество от избрани обекти и осигурява обратната връзка от потребителския интерфейс към базата от данни при изтриване на обекти от екрана. След като се получат двумерните координати на образите на обектите от FastMap алгоритъма, получения образ се „раздува" до размера на работното поле за визуализацията. Така във всеки един момент обектите се показват максимално скалирани, така че да се побират в наличното пространство. Визуализацията на обектите е реализирана чрез запълнени кръгчета с цвета на съответния клъстер. Радиусът на кръгчетата може да се регулира от опциите. Степента на Alpha-Blending ефекта също може да се регулира. Непрекъснато се следи позицията на курсора и при доближаване на достатъчно малко разстояние до някой обект, той автоматично се избира, извлича се от базата неговата информация и се показва в левия панел.

*Database*
В този клас е събрана функционалността по отваряне и записване на входните файлове, прочитане на данните, извличане на метаданните (атрибутите на примерите и техния клас), зареждане и визуализиране на данните в таблици, преобразуване на данните до по-бърз за използване вид (индексиране на номиналните стойности и Z-нормализиране). Този модул е отговорен и за допълнителната обработка на данните за извличането на различните статистики.

*Metadata*
Тук се поддържа информацията за начина, по който да се прочетат/запишат данните, за атрибутите и класа на примерите. За всеки номинален атрибут се поддържа списък от възможни стойности. Класът се третира като специален вид номинален атрибут. За всеки атрибут и за всяка номинална стойност се пазят временни статистически данни, които се използват при нормализацията и при извличането на статистики. При зареждане на извадка от .data файл, програмата търси файл със същото име като файла с данните, но с разширение .names.

*Attribute / NominalAttribute / ContinuousAttribute*
Attribute е абстрактен клас, който определя общата функционалност за всички атрибути. Останалите два класа наследяват Attribute. Така се позволява на *Metadata да работи полиморфно* със списък от атрибути с произволен тип. При зададени конкретни данни, обектите от тип ContinuousAttribute поддържат статистическа информация за средната стойност на атрибута и стандартното отклонение. Съответно, обектите от тип NominalAttribute поддържат информация за разпределението на данните по всяка номинална стойност, както и за диаметрите на клъстерите, при условие, че атрибута указва номер на клъстер. Тези статистики се изчисляват от обекта Database, при поискване от потребителя.

# Възможности за бъдещо развитие

В съответствие с целта на дипломната работа, успешно бяха създадени два инструмента за конструиране и визуализиране на извадки от медицински бази от данни.

В уводната част направихме кратко въведение в предметната област и показахме необходимостта от визуализация на голям обем многомерни медицински данни. Описахме задачите на дисциплината ИЗД и уточнихме мястото, което настоящата разработка заема в него. В края на въведението дефинирахме точно и ясно целта на дипломната работа.

В хода на проектирането направихме анализ на изискванията към Конструктора и проектирахме конкретни стъпки за създаване на извадки от големи бази от данни. След това направихме описание на алгоритъма FastMap за намаляване размерността на данни и на негова база проектирахме инструмента Визуализатор.

В частта за реализация обосновахме избора на средства за написване на двата инструмента, описахме реализирания потребителски интерфейс и посочихме функционалността на основните класове.

Текущата реализация на Конструктора е напълно функционираща и чрез нея с лекота могат да се създават сложни SQL заявки. Реализирани са абсолютно всички описани в изложението стъпки, както и някои допълнителни, създадени за по-лесно приспособяване на инструмента към схемата на базата от данни. За доработването на функционалността и изчистването на потребителския интерфейс съм задължен на хилядите абонати на медицински софтуер „Хипократ". Това е може би най-разпространеният медицински софтуер в България, с който ежедневно работят над 2000 общопрактикуващи лекари и специалисти. Конструкторът успешно беше интегриран в програма „Хипократ" и вече повече от 1 година потребителите на програмата го използват за създаване на гъвкави извадки от своите бази с медицински данни.

Благодарение на отправените от потребителите предложения, критики и препоръки към Конструктора той има толкова завършен вид в момента. Извън границите на проектираната функционалност, към Конструктора бяха добавени допълнителни функции за улеснение на потребителите, като например възможността за записване на конструирани условия и повторното им готово използване под формата на „Готови справки". Тази функционалност е една от най-приятните за потребителите възможности за пестене на време в ежедневната им работа.

Също така, един от най-големите медицински центрове в България (ДКЦ-2 в Плевен), който работи с програма „Хипократ", предостави огромната си по обем база с медицинска информация за тестване на двата инструмента и допълнителното им оптимизиране за работа с голям обем данни. Точно тези тестове доведоха до идеята за допълнителна обработка и индексиране на данните преди пускане на FastMap алгоритъма. Разбира се, личната информация от тяхната база беше старателно премахната, за да не могат да се идентифицират отделни пациенти.

Текущата реализация на Визуализатора също е напълно функционираща и с нея удобно се визуализират не само медицински извадки, но и произволни извадки с данни. Допълнително реализираните операции за манипулиране на обектите от визуализацията придават изключително полезна интерактивност на инструмента и му помагат наистина да се превърне във визуално средство за извличане на закономерности от данните.

За разлика от Конструктора, Визуализаторът не е чак толкова добре изтестван от потребители и това определя до голяма степен по-слабото му развитие извън границите на проектираната функционалност. Основен недостатък в текущата реализация е липсата на адекватна обработка на липсващите стойности в извадката. В момента програмата не обработва липсващите стойности на атрибутите, а просто пропуска редовете с тях. Добре би било да се реализират някои от описаните при проектиране на Визуализатора подходи за справяне с липсващите стойности. Това би имало голямо значение най-вече за такива извадки, в които процентът на редовете с липсващи стойности е значителен.

Друга възможност за подобряване на текущата реализация е добавяне на още метрики за разстояния. В момента напълно реализирани са функциите за разстояние между номинални и непрекъснати атрибути. Това на практика е достатъчно за визуализация на всяка произволна извадка, но не може да осигури достатъчно добри резултати при визуализиране например на наредени номинални атрибути. Например номиналните стойности {„малко", „средно", „голямо"} ще се считат за еднакво раздалечени една от друга точно както и номиналните стойности {„бяло", „зелено", „червено"}, въпреки че за първите съществува наредба, според която "малко" трябва да е по-близо до „средно", отколкото до „голямо". В подобни случаи е удачно да се използват методи от размитата логика (fuzzy logic), които да определят сходството на една величина със стойности от типа на „малко" и „голямо".

По време на разработването на Визуализатора възникна идея за създаване на специфична мярка за разстояние между медицински диагнози. Наскоро в България беше въведена в употреба 10-та ревизия на Международната Класификация на Болестите (МКБ-10). Това представлява една унифицирана световна номенклатура, която съпоставя на всяка болест един уникален идентификационен код. Пример за такъв код е „М51", който съответства на „Болест на междупрешленните дискове". За максимална полза от Визуализатора при работа с такива диагнози, добре би било да се разработи специална функция за разстояние между две диагнози, кодирани по МКБ-10. Така програмата ще знае, че М51 е по-близо до болестта „Ревматоиден артрит", отколкото до болестта „Психично разстройство". Това, разбира се, може да стане само в тясно сътрудничество с медицински експерти и за съжаление не е реализирано в момента.

Едно ограничение на текущата реализация на Визуализатора е фактът, че той може да показва само двумерни представяния на данните. Този недостатък не е наложен от FastMap алгоритъма, а по-скоро от използваната библиотека Drawing2D за визуализация. Едно възможно бъдещо разширение на Визуализатора би било в посока визуализация на тримерни данни чрез използване на платформа като DirectX или OpenGL. Това, разбира се, ще доведе до множество проблеми с начина на придвижване в 3D пространството и начина на избиране на обекти, което за момента доста удобно е реализирано в 2D пространството.

Може да се мисли в бъдеще и за интеграция на двата инструмента с други програми за ИЗД. Реализираният файлов формат за обмен на данни и мета-данни осигурява добра възможност за това. Например, възможно е да се комбинира Конструкторът с инструмент за клъстеризиране на медицински данни, които след това да бъдат визуализирани и да се оцени качеството на клъстеризация от статистическите данни във Визуализатора. Пример за подобен инструмент за клъстеризиране е така нареченият DRG групер (Diagnostically Related Groups), който НЗОК (Националната Здравно-осигурителна Каса на България) използва за групиране на диагнозите при определяне на държавното подпомагане към болниците за различните клинични пътеки.

# Литература

## *Приложение 1. Списък на възприетите преводи на английски термини*

В тази таблица са събрани всичките възприети в дипломната работа преводи от английски термини на български език.

| Английски термин | Възприет превод на български език |
|---|---|
| Computer application | Компютърно приложение |
| Context-sensitive | Контекстно-зависим |
| Database | База от данни |
| Data Mining | Извличане на закономерности от данни |
| Delimited text file | Текстов файл с разделител на данните |
| Framework | Платформа |
| Fuzzy logic | Размита логика |
| Implementation | Реализация |
| Instance | Екземпляр |
| Logical development | Логическо развиване |
| Relational Database Management System | Система за управление на бази от данни |
| Refinement | Уточняване |
| Structured Query Language | Език на структурирани заявки |
| Tool | Средство, инструмент |
| User-friendly | Удобна за потребителя/използване |
| Workflow | Работна последователност |

## *Приложение 2. Списък на използваните съкращения*

В тази таблица са събрани всички съкращения, използвани в изложението. Някои от тях са добре утвърдени сред компютърните специалисти, други са въведени в настоящата работа с цел улесняване на записа.

| Съкращение | Пълно наименование |
|---|---|
| ИЗД | Извличане на закономерности от данни |
| МКБ-10 | Международната Класификация на Болестите, 10-та ревизия |
| НЗОК | Националната Здравно-осигурителна Каса на България |
| СУБД | Система за управление на бази от данни (вж. RDBMS) |
| DRG | Diagnostically Related Groups |
| DTD | XML Document Type Definition |
| MVDM | Modified Value Difference Metric |
| RDB | Relational Database |
| RDBMS | Relational Database Management System |
| SQL | Structured Query Language |
| XML | Extensible Markup Language |
| XSD | XML Schema Definition |